\documentclass[a4paper,fleqn,sort&compress,figures]{cas-sc}

\usepackage[numbers]{natbib}
\usepackage{subfigure}

\begin{document}
\let\WriteBookmarks\relax
\def\floatpagepagefraction{1}
\def\textpagefraction{.001}
\shorttitle{Anisotropic $q$-state clock model}
\shortauthors{M. \v{Z}ukovi\v{c}}

\title [mode = title]{Competing anisotropies and phase transitions in the $q$-state clock model with a $p$-fold crystalline field}                      
%
%

\author{M. \v{Z}ukovi\v{c}}[orcid=0000-0001-6241-299X]
\ead{milan.zukovic@upjs.sk}

\credit{Conceptualization, Investigation, Methodology, Software, Data curation, Visualization, Formal analysis, Validation, Writing - original draft, Writing - review \& editing, Resources, Project administration}

\affiliation[1]{organization={Department of Theoretical Physics and Astrophysics, Institute of Physics, Faculty of Science, Pavol Jozef \v{S}af\'arik University in Ko\v{s}ice},
                addressline={Park Angelinum 9}, 
                city={Ko\v{s}ice},
                postcode={04154}, 
                country={Slovakia}}

\begin{abstract}
We study the two-dimensional $q$-state clock model in the presence of an additional $p$-fold symmetry-breaking crystalline field using Monte Carlo simulations. While the pure clock model exhibits Berezinskii--Kosterlitz--Thouless (BKT) transitions for sufficiently large $q$, the effect of competing discrete anisotropies on this topological phase remains nontrivial. We show that even weak crystalline fields qualitatively modify the phase diagram by suppressing the BKT phase and inducing transitions to states with true long-range order. The resulting behavior depends sensitively on the interplay between the intrinsic $\mathbb{Z}_q$ symmetry and the imposed $\mathbb{Z}_p$ anisotropy. In particular, in the six-state clock model for $p=2$ we observe qualitatively different scenarios depending on the sign of the field: a single transition for $h_2>0$ and a two-step ordering process for $h_2<0$ with an intermediate ordered phase. For $p=3$, the system exhibits a direct transition consistent with three-state Potts criticality. These results demonstrate that the phase structure cannot be inferred from symmetry considerations alone, but is governed by the competition between distinct locking mechanisms. Our findings provide a discrete counterpart to the multi-frequency sine-Gordon description of generalized $XY$ models and illustrate how additional anisotropies reshape topological phase transitions in two dimensions.
\end{abstract}


\begin{highlights}
\item Competing anisotropies suppress BKT criticality in the clock model
\item Crystalline fields induce nontrivial ordering sequences in 2D
\item Two-step ordering emerges for negative twofold crystalline fields
\item Phase structure depends on competition between $\mathbb{Z}_q$ and $\mathbb{Z}_p$ symmetries
\item Monte Carlo results reveal crossover beyond symmetry arguments
\end{highlights}


\ExplSyntaxOn
\keys_set:nn { stm / mktitle } { nologo }
\ExplSyntaxOff

\maketitle

\section{\label{sec:intro}Introduction}
Low-dimensional systems with continuous or discrete symmetries provide a fertile ground for studying unconventional phase transitions driven by topological excitations and competing interactions. A paradigmatic example is the two-dimensional ($2$D) $XY$ model, which undergoes a Berezinskii--Kosterlitz--Thouless (BKT) transition associated with the unbinding of vortex--antivortex pairs \cite{Berezinskii1972,KosterlitzThouless1973,Kosterlitz1974}. The introduction of symmetry-breaking perturbations, such as discrete anisotropies, can qualitatively modify this behavior and lead to a rich variety of critical phenomena \cite{Jose1977}.

The $q$-state clock model provides a natural framework to study such effects, as it interpolates between discrete and continuous symmetries. For small values of $q$, the clock model exhibits conventional symmetry-breaking phase transitions. The cases $q=2$, $3$, and $4$ correspond to the Ising, three-state Potts, and four-state Potts universality classes, respectively, and display true long-range order (LRO) below a critical temperature $T_c$~\cite{Wu1982,Baxter1982}. On the other hand, for sufficiently large $q \geq 5$, the model exhibits two BKT transitions with an intermediate critical phase characterized by quasi-long-range order (QLRO) and true LRO at low temperatures \cite{Elitzur1979,Lapilli2006}, albeit a recent higher-order mean-field study argues that the lower temperature transition is of a large-order spontaneous symmetry-breaking type~\cite{Goswami2025}. 

The precise nature of the transitions for intermediate values of $q$, particularly for $q=5$ and $q=6$, has been the subject of extensive numerical investigation. More recent studies employing large-scale Monte Carlo simulations, helicity modulus analysis, conformal invariance, corner-transfer matrix renormalization group, and tensor-network techniques~\cite{Tomita2002,Matsuo2006,Baek2013,Kumano2013,Li2020,Ueda2020,polackova2023,Shi2026} have largely agreed on the two-transition scenario, while emphasizing the presence of strong finite-size effects and slow crossover behavior near the lower transition. Most studies focused on a square-lattice implementation but some studies also considered other lattice geometries~\cite{otsuka2023,okabe2025}.

In the critical phase of these models, the discrete $\mathbb{Z}_q$ anisotropy becomes irrelevant at long wavelengths, leading to an emergent $U(1)$ symmetry. However, additional perturbations may destabilize it and induce conventional ordering. A particularly interesting class of perturbations is provided by crystalline fields that explicitly break rotational symmetry by favoring a subset of spin orientations. In the context of generalized $XY$ models, such terms correspond to additional harmonics in a sine-Gordon description, as analyzed by Jos\'{e} \textit{et al.} \cite{Jose1977}. In that framework, the relevance of competing cosine terms depends on their scaling dimensions at the BKT fixed point, leading to a rich phase structure with multiple transitions and crossover phenomena.

In the present work, we consider the discrete analogue of this problem by studying the $q$-state clock model on a square lattice in the presence of a $p$-fold symmetry-breaking crystalline field $h_p$. The Hamiltonian takes the form
\begin{equation}
\mathcal{H} = -J \sum_{\langle ij \rangle} \cos(\theta_i - \theta_j)- h_p \sum_i \cos(p \theta_i),
\label{Hamilt_aniso}
\end{equation}
where $J>0$ denotes a ferromagnetic nearest-neighbor coupling, $\theta_i = 2\pi n_i/q$ and $n_i = 0,1,\dots,q-1$. Unlike in the continuous case, where anisotropies can be treated perturbatively, the clock model corresponds to the strong-anisotropy limit in which angular variables are already restricted to discrete values. The addition of a second discrete symmetry therefore introduces a competition between distinct locking mechanisms, whose interplay is not determined by symmetry alone.

From a symmetry perspective, one can identify the degeneracy of the ground states and the associated residual symmetry groups. However, this information is generally insufficient to determine the structure of the phase diagram. In particular, it does not specify whether ordering occurs in a single step or through multiple transitions, whether intermediate phases are present, or how the BKT phase evolves under finite perturbations. These questions depend on the detailed interplay between domain walls, vortex excitations, and entropy, and require a nonperturbative analysis.

Here we address these issues using large-scale Monte Carlo simulations. We focus on representative cases with $q=6$ and $p=2,3$, which are demonstrated to exhibit a variety of behaviors depending on the sign and strength of the crystalline field. We note that the case $p=3$ does not represent lattice-generated but rather an externally imposed anisotropy and such a choice is motivated as a generic competing-symmetry perturbation. We show that in the presence of the crystalline field the BKT behavior is replaced, within the accessible system sizes, by transitions consistent with conventional LRO. Moreover, we find that the ordering process depends sensitively on the structure of the competing anisotropies, leading in some cases to multiple transitions and intermediate phases that are not fixed by symmetry arguments.

Our results provide a discrete realization of the competition between multiple cosine terms in generalized $XY$ models and illustrate how additional anisotropies reshape topological phase transitions. They also highlight the importance of nonperturbative effects in determining phase diagrams of systems with competing discrete symmetries.

The central question addressed in this work is therefore not merely the symmetry of the resulting ordered phases, but rather how competing discrete anisotropies modify the topological ordering mechanisms of the pure clock model. In particular, we ask whether the BKT phase survives finite crystalline fields, whether ordering occurs in one or multiple steps, and how these phenomena depend on the interplay between intrinsic and externally imposed anisotropies.

\section{\label{sec:method}Method}
To investigate the phase transition characteristics of the current spin system, we employed Monte Carlo (MC) simulations utilizing the Metropolis algorithm for stochastic state updates. Calculations were performed on lattices of size $L^{2}$ with linear dimensions $L=24-120$, under the imposition of periodic boundary conditions. Thermodynamic observables were evaluated as a function of the reduced temperature $k_{B}T/J$. To fix the temperature scale, in the following we set $J = k_B = 1$. For each data point, the system was evolved for a total of $N=5 \times 10^{5}$ Monte Carlo sweeps (MCS), after discarding the initial $N_{0}=0.2 \times N$ sweeps to ensure sufficient thermalization. Simulations were initiated at high temperatures from random configurations and subsequently cooled in increments of a typical value $\Delta T=0.025$. By utilizing the final microstate of the preceding temperature step as the initial configuration for the next, we minimized thermalization latency and ensured the system remained near quasi-equilibrium throughout the cooling protocol.

To determine the critical exponents, we perform a finite-size scaling (FSS) analysis, augmented by reweighting techniques~\cite{ferrenberg1988new,ferrenberg1989new}. For these calculations, the simulation length is extended to $N = 10^7$ MCS. Such prolonged observation times are imperative due to the critical slowing down observed in the system. Statistical uncertainties are rigorously evaluated using the $\Gamma$-method~\cite{wolff2004monte}. Unlike conventional binning techniques that account for autocorrelations only implicitly, the $\Gamma$-method provides a more robust error estimation by explicitly computing the relevant autocorrelation functions. Our analysis confirms that the selected simulation parameters, combined with the $\Gamma$-method, maintain statistical errors within acceptable confidence intervals for all investigated lattice sizes.

We measure and analyze the specific heat per spin $c$
\begin{equation}
c=\frac{\langle {\mathcal H}^{2} \rangle - \langle {\mathcal H} \rangle^{2}}{L^2T^{2}},
\label{c}
\end{equation}
the generalized magnetizations $m_k$, $k=1,2,\dots,q$,
\begin{equation}
m_k=\langle M_{k} \rangle/L^2=\left\langle\Big|\sum_{j}\exp(ik\phi_j)\Big|\right\rangle/L^2,
\label{mk}
\end{equation}
and the corresponding susceptibilities $\chi_{k}$
\begin{equation}
\label{chi_mk}\chi_{k} = \frac{\langle M_{k}^{2} \rangle - \langle M_{k} \rangle^{2}}{L^2T},
\end{equation}
which in the presence of the symmetry-breaking fields can be useful as order parameters ($m_k$) for identifying different emerging phases and the corresponding response functions ($\chi_k$) for establishing the nature of the respective phase transitions. Furthermore, we calculate the derivatives of the generalized magnetizations
\begin{equation}
\label{dm}dm_{k} = \frac{\partial}{\partial \beta}\langle M_{k} \rangle = \langle M_{k} {\mathcal H}\rangle -\langle M_{k} \rangle\langle {\mathcal H} \rangle,
\end{equation}
and the derivatives of the logarithms of their first two moments
\begin{equation}
\label{dlm}dlm^l_{k} = \frac{\partial}{\partial \beta}\ln\langle M_{k}^l \rangle = \frac{\langle M_{k}^l {\mathcal H}\rangle}{\langle M_{k}^l \rangle}- \langle {\mathcal H} \rangle,\ l=1,2.
\end{equation}

At second-order phase transitions the maxima of the above quantities scale with the lattice size as
\begin{equation}
\label{fss_c}c_{max}(L) \propto L^{\alpha/\nu},
\end{equation}
\begin{equation}
\label{fss_chi}\chi_{k,max}(L) \propto L^{\gamma/\nu},
\end{equation}
\begin{equation}
\label{fss_dmk}dm_{k,max}(L) \propto L^{(1-\beta)/\nu},
\end{equation}
\begin{equation}
\label{fss_dlmk}dlm^{l}_{k,max}(L) \propto L^{1/\nu},
\end{equation}
where $\alpha,\beta,\gamma$ and $\nu$ represent the critical exponents of the specific heat, the order parameter $m_k$, the susceptibility $\chi_k$, and the correlation length, respectively. 

\section{\label{sec:res}Results}
\subsection{\label{subsec:res_iso}Isotropic case}

In the present study we will focus on the effects of the crystalline fields in the well-studied clock model with $q=6$. The critical behavior of the isotropic six-state clock model is rather well-known. It is generally accepted that there are two phase transitions (both most likely of the BKT nature) that separate three phases: a LRO ferromagnetic phase at low temperature, a disordered paramagnetic phase at high temperature and an intermediate phase that is predicted to be a quasiliquid BKT phase.

\begin{figure}[t!]
\centering
\subfigure{\includegraphics[scale=0.4,clip]{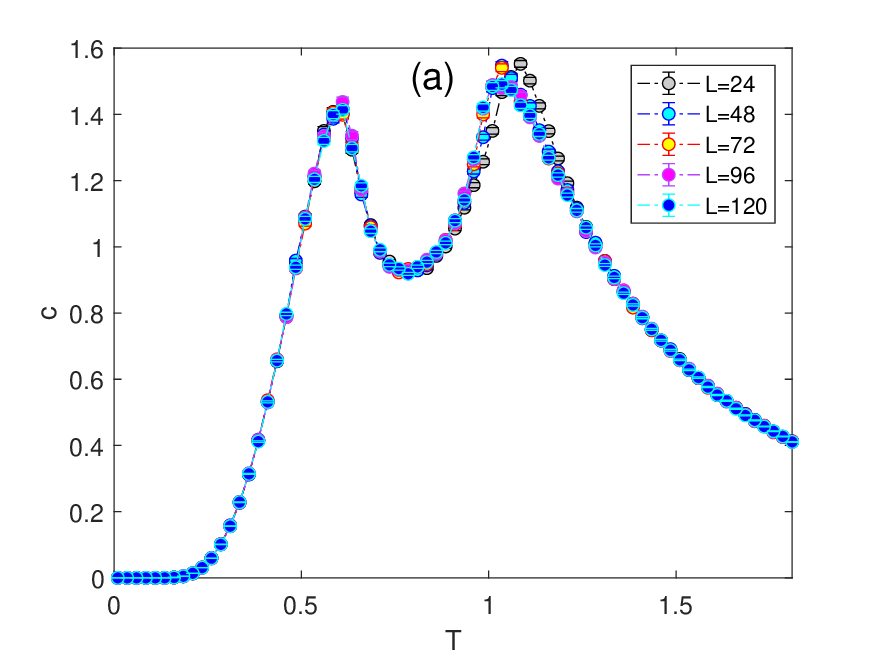}\label{fig:c-T_D0_L24-120}}\hspace*{-5mm}
\subfigure{\includegraphics[scale=0.4,clip]{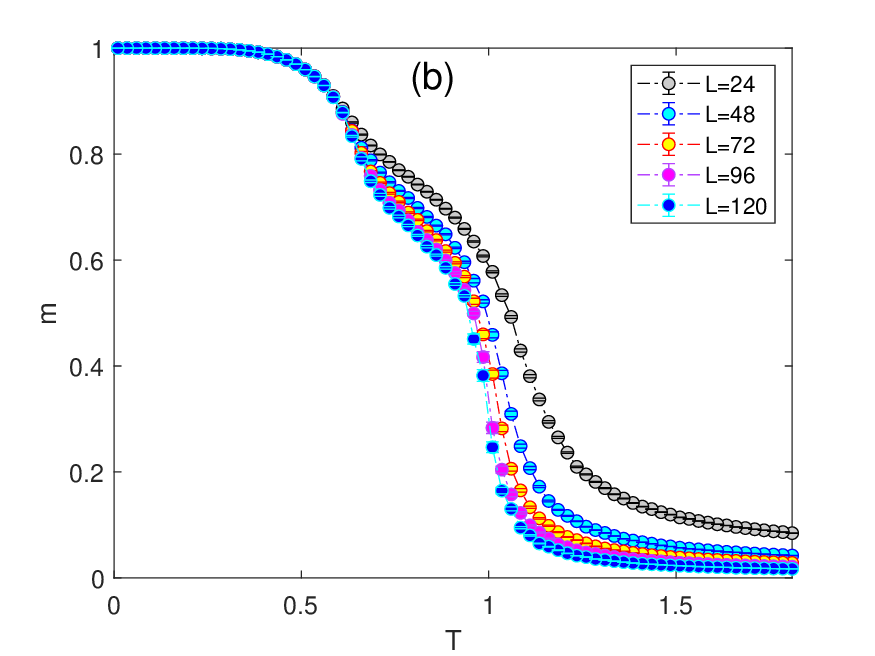}\label{fig:m-T_D0_L24-120}}\hspace*{-5mm}
\subfigure{\includegraphics[scale=0.4,clip]{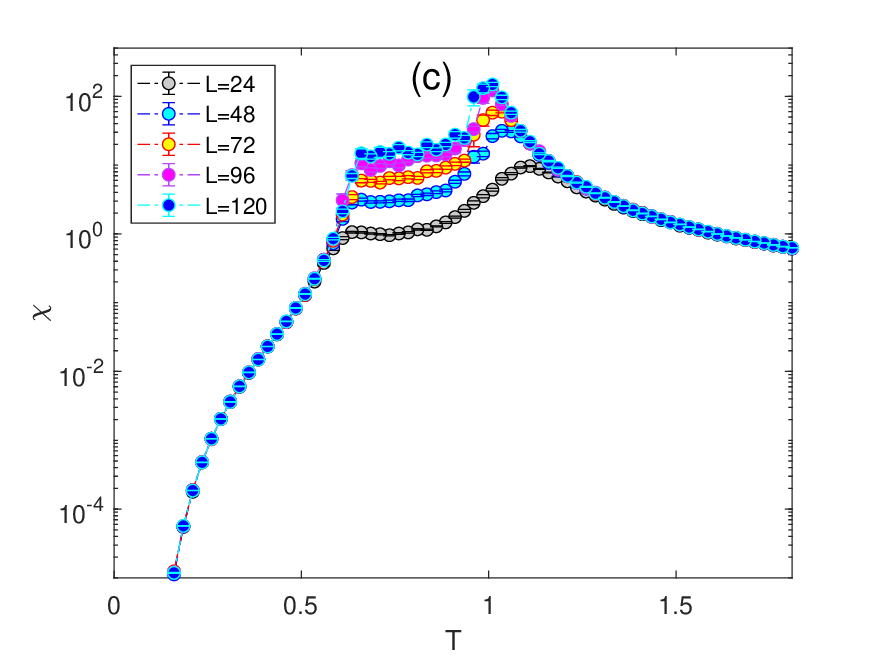}\label{fig:chi-T_D0_L24-120}}
\caption{Temperature dependencies of (a) the specific heat, (b) the magnetization and (c) the magnetic susceptibility of the isotropic six-state clock model, for various lattice sizes.}
\label{fig:x-T_D0}
\end{figure}

Before including effects of the crystalline fields it would be instructive to present thermodynamic behavior of the relevant studied quantities in the isotropic case. In Fig.~\ref{fig:x-T_D0} we show temperature dependencies of the specific heat, magnetization and magnetic susceptibility (correspond to $m_1$ and $\chi_1$ according to Eqs.~(\ref{mk}) and (\ref{chi_mk})), for various lattice sizes. The presence of the expected two phase phase transitions is reflected in the two round peaks in the specific heat (at $T_{c1} \approx 0.6$ and $T_{c2} \approx 1$)\footnote{The list of more precise values obtained by different approaches can be found e.g., in Ref.~\cite{Ueda2020}.}, which show little sensitivity to the system size. The region between $T_{c1}$ and $T_{c2}$ is characterized by the emergent U(1) symmetric BKT phase with the power-law decay of the magnetization and divergence of the susceptibility with the increase of the lattice size. The behavior of the latter two quantities in this critical phase is controlled by the temperature-dependent critical exponent $\eta(T)$.

\subsection{Effect of the crystalline field: competing anisotropies}

To gain insight into the role of the crystalline field, it is useful to first examine the energies of the possible spin orientations (Table~\ref{table:energy}). For a given value of $p$ and field $h_p$, the single-site contribution $-h_p \cos(p\theta)$ selects a subset of angles that minimize the energy. Combined with the intrinsic $\mathbb{Z}_q$ discretization, this leads to a set of energetically favored states (in bold for negative and in roman for positive values of $h_p$) whose degeneracy depends on $p$ and the sign of $h_p$.

\begin{table}[t!]
\caption{Energy spectra of the six-state clock model for the crystalline field terms $h_p$ corresponding to different spin states $n_i=0,1,\dots,5$, and different types of the symmetries $p=1,2,\dots,5$.} 
\label{table:energy}
\begin{tabular}{l|cccccc}
\hline
\hfil $n_i$ & \hfil   0 & \hfil 1  & \hfil  2 & \hfil 3 & \hfil  4  & \hfil 5 \\ \hline
  $p=1$ & $\mathrm{-h_1}$ &   $-h_1/2$ &    $h_1/2$ &   $\mathbf{h_1}$&    $h_1/2$&       $-h_1/2$\\
  $p=2$ & $\mathrm{-h_2}$ &   $\mathbf{h_2/2}$ &   $\mathbf{h_2/2}$ &   $\mathrm{-h_2}$&    $\mathbf{h_2/2}$ &    $\mathbf{h_2/2}$\\
  $p=3$ & $\mathrm{-h_3}$ &   $\mathbf{h_3}$ &    $\mathrm{-h_3}$ &   $\mathbf{h_3}$&    $\mathrm{-h_3}$&       $\mathbf{h_3}$\\
	$p=4$ & $\mathrm{-h_4}$ &   $\mathbf{h_4/2}$ &    $\mathbf{h_4/2}$ &   $\mathrm{-h_4}$&    $\mathbf{h_4/2}$&   $\mathbf{h_4/2}$\\
	$p=5$ & $\mathrm{-h_5}$ &   $-h_5/2$ &    $h_5/2$ &   $\mathbf{h_5}$&    $h_5/2$&       $-h_5/2$\\
\hline
\end{tabular}
\end{table}

While this analysis identifies the low-energy manifold and its symmetry, it does not fully determine the phase behavior. In particular, it does not specify whether the system undergoes a single transition or multiple transitions upon heating, whether intermediate phases are present, or how the BKT phase of the pure clock model evolves under finite $h_p$. These questions depend on the interplay between competing anisotropies, domain-wall excitations, and topological defects, and therefore require a detailed numerical investigation.

From a broader perspective, the present model can be viewed as the strong-anisotropy limit of a generalized $XY$ model with multiple cosine terms. In the renormalization-group analysis of Jos\'{e} \textit{et al.} \cite{Jose1977}, such terms correspond to competing locking potentials whose relevance depends on their scaling dimensions. In the discrete clock model, however, the intrinsic $\mathbb{Z}_q$ anisotropy is already strong, and the additional $\mathbb{Z}_p$ field introduces a second locking mechanism. The resulting behavior is therefore governed by the competition between these terms rather than by symmetry alone.

Let us first address the crystalline fields that will have either trivial or the same effects on the model's critical behavior. The field $h_1$ corresponds to an external magnetic field, which explicitly breaks the global $\mathbb{Z}_q$ symmetry and thus no thermodynamic phase transition can occur. Namely, it pins one direction and destroys both BKT transitions. As evidenced from Table~\ref{table:energy}, the preferred states are 0 ($h_1>0$) or 3 ($h_1<0$) that correspond to positive and negative directions, respectively, along the $x$ axis. From Table~\ref{table:energy} one can also see that energetic contributions from the $h_4$ field are exactly the same as for $h_2$ and thus one can expect the same thermodynamic as well as critical behavior for both $h_2$ and $h_4$. Similar statement also holds for the cases $h_1$ and $h_5$ and thus the field $h_5$ will have the same effect as an external magnetic field. Thus, below we will focus on the representative (non-trivial) cases of $p=2$ and $3$.

We remark that some choices of the crystalline field, such as $p=3$, are not dictated by the symmetry of the underlying square lattice and should be viewed as an externally imposed anisotropy rather than an intrinsic lattice effect. In this sense, the model represents a more general class of systems in which discrete degrees of freedom are subject to competing symmetry-breaking fields. Similar situations arise in a variety of physical contexts, including adsorbed monolayers on substrates with different symmetry, Josephson-junction arrays with engineered anisotropies, and artificial spin systems. From a theoretical perspective, the $p=3$ case provides a simple example of competing discrete symmetries that allows one to probe the general mechanisms discussed in this work. Our results indicate that the resulting behavior is governed by the interplay between these symmetries rather than by the lattice structure alone. We now turn to the Monte Carlo results, which reveal several distinct scenarios depending on $p$ and the sign of $h_p$.

\subsubsection{Case $p=2$}
\begin{figure}[t!]
\centering
\subfigure{\includegraphics[scale=0.5,clip]{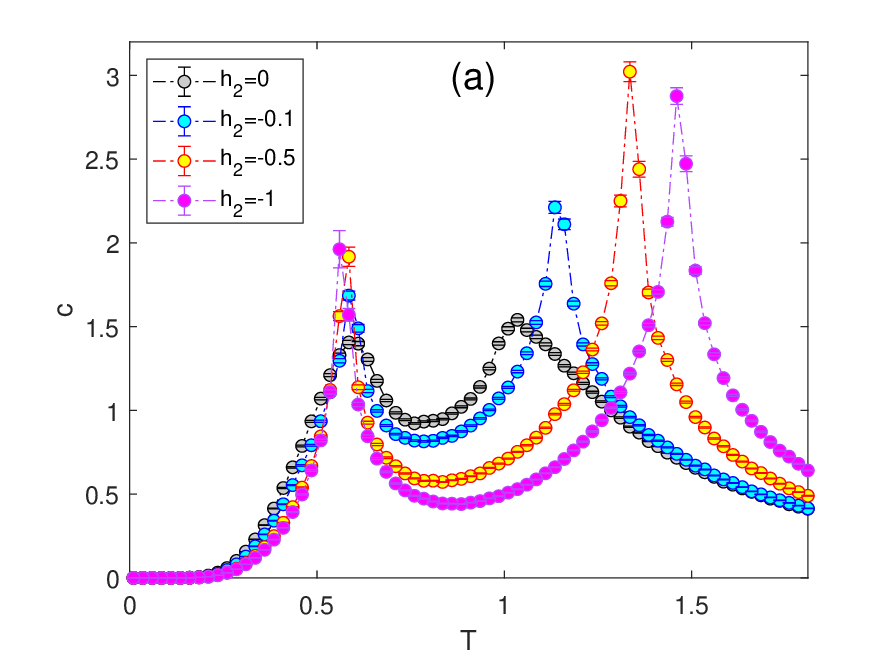}\label{fig:c-T_p2_D_neg_L72}}
\subfigure{\includegraphics[scale=0.5,clip]{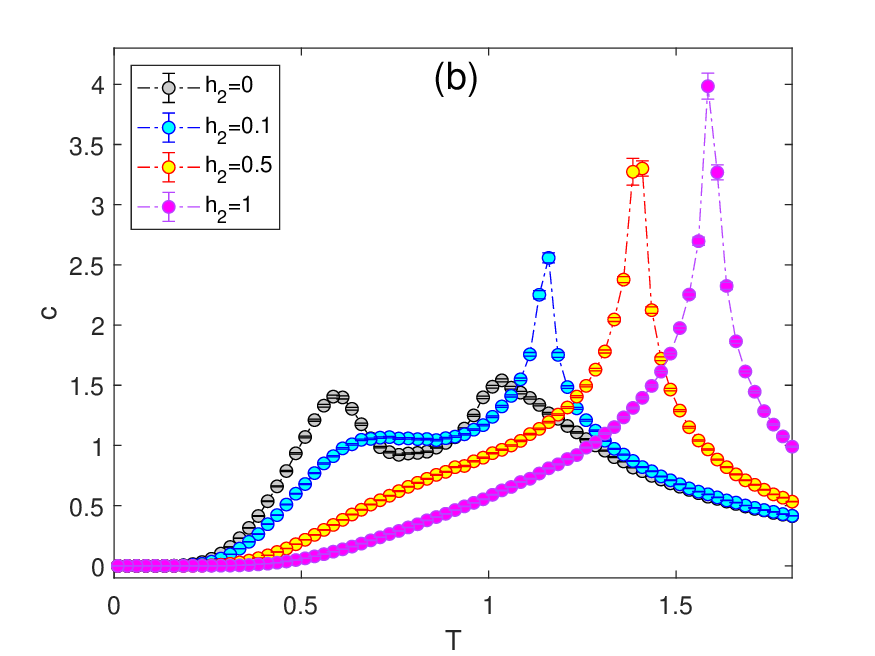}\label{fig:c-T_p2_D_pos_L72}}
\caption{Temperature dependencies of the specific heat for $p=2$, $L=72$, and several (a) negative and (b) positive values of $h_2$ ($h_2=0$ case is also included for reference).}
\label{fig:c-T_p2_L72}
\end{figure}

Figure~\ref{fig:c-T_p2_L72} demonstrates effects of the field $h_2$ on the specific heat behavior. For $h_2>0$, the crystalline field favors two orientations separated by $\pi$, effectively selecting a subset of the clock states. The simulations show a single phase transition from the disordered phase to a symmetry-broken ordered phase. This is evident from the specific heat presented in Fig.~\ref{fig:c-T_p2_D_pos_L72}, which shows a single sharp peak that moves toward higher temperatures with the increasing $h_2$. This peak evolves from the high-temperature BKT transition at $h_2=0$, while the peak associated with the low-temperature transition transforms to a hump for small $h_2$ and tends to disappear as $h_2$ increases.

For $h_2<0$, the situation is qualitatively different. In this case, the energetically favored states form a larger set, allowing for a two-step ordering process. The simulations clearly resolve two distinct transitions upon cooling: a high-temperature transition from the disordered phase to an intermediate ordered phase, followed by a second transition into the fully ordered ground state (Fig.~\ref{fig:c-T_p2_D_neg_L72}). Thus, the two round peaks in the specific heat, observed in the isotropic case, transform into two sharp peaks. The position of the low-temperature peak is little sensitive to the value of $h_2$ but the high-temperature one is clearly shifted to higher temperatures with the increasing intensity of $h_2$. As will be demonstrated below, the intermediate phase is characterized by partial symmetry breaking and LRO within a subset of states. The existence of two separate transitions is not fixed by symmetry alone and reflects the presence of competing ordering tendencies induced by the crystalline field.

\begin{figure}[t!]
\centering
\subfigure{\includegraphics[scale=0.4,clip]{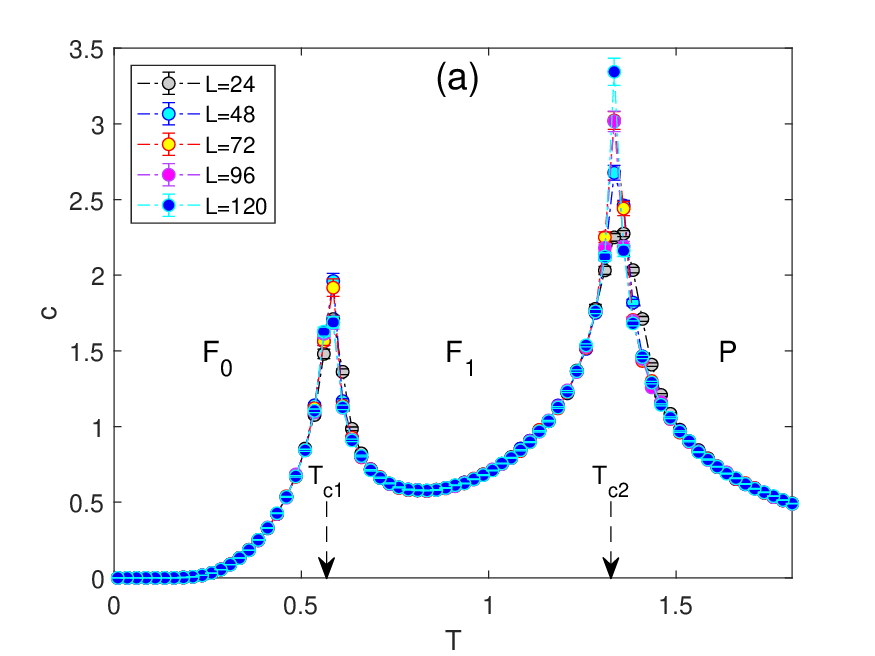}\label{fig:c-T_p2_D_-0_5_L24-120}}\hspace*{-5mm}
\subfigure{\includegraphics[scale=0.4,clip]{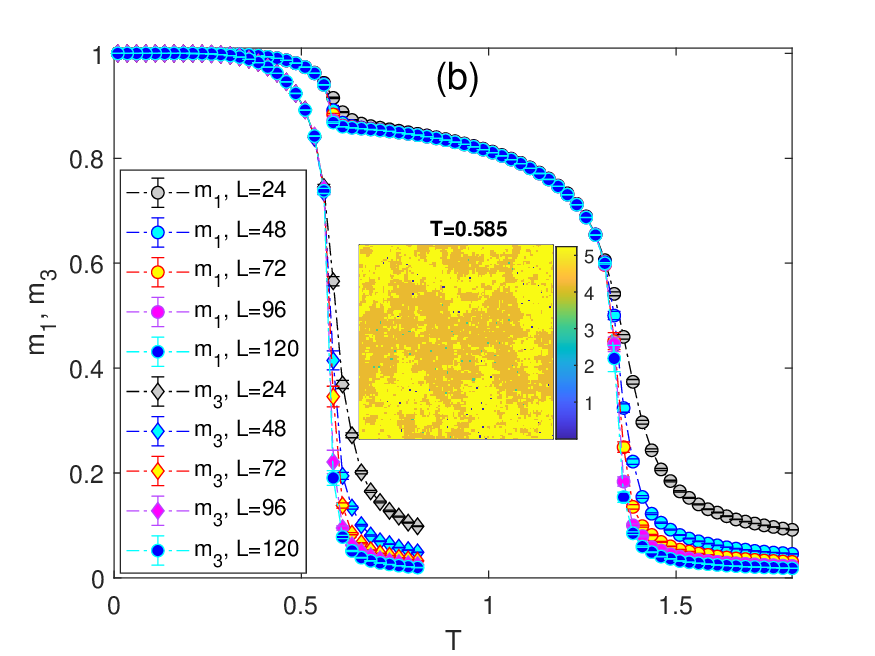}\label{fig:m-T_p2_D_-0_5_L24-120}}\hspace*{-5mm}
\subfigure{\includegraphics[scale=0.4,clip]{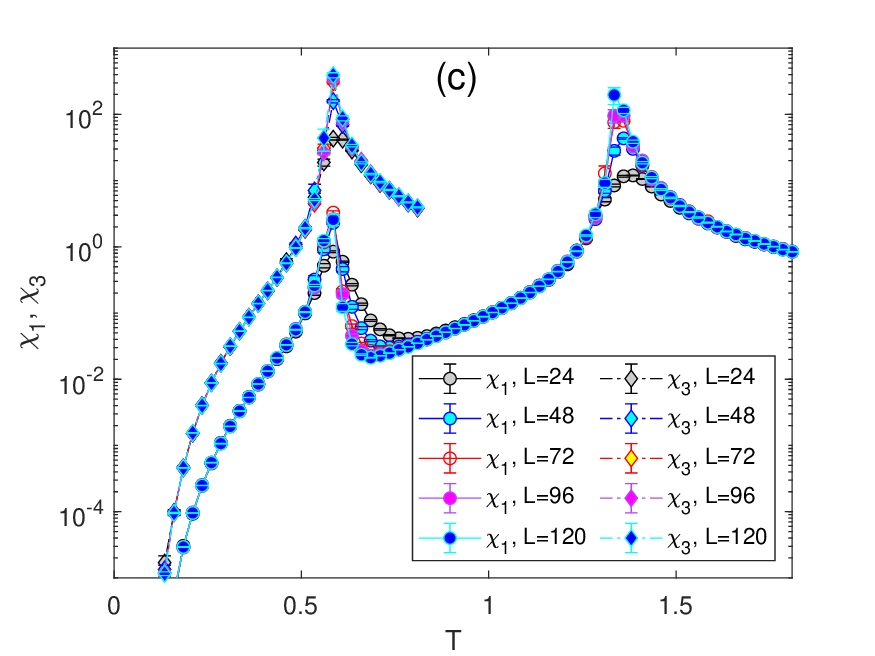}\label{fig:chi-T_p2_D_-0_5_L24-120}}\\ \vspace*{-3mm}
\subfigure{\includegraphics[scale=0.4,clip]{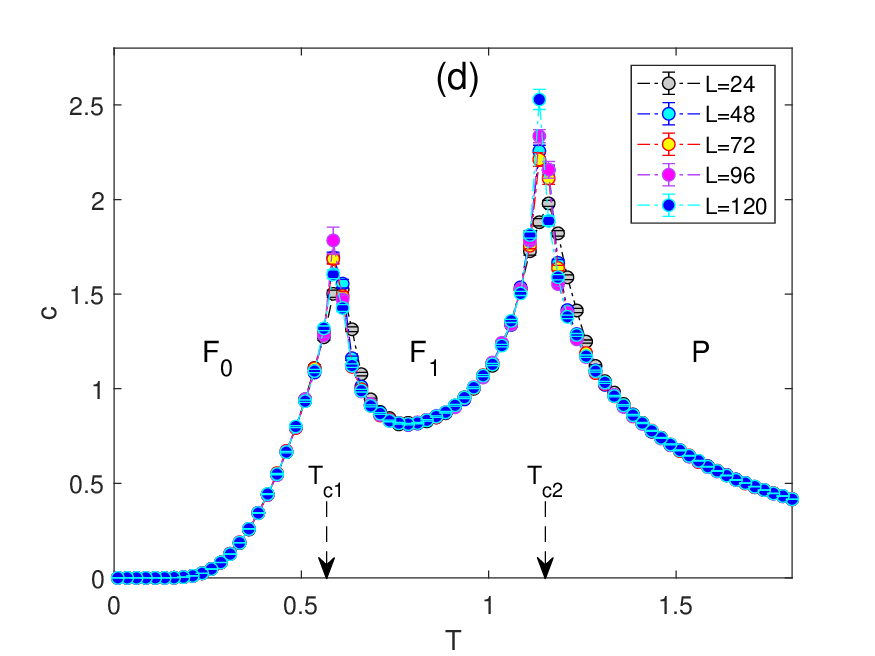}\label{fig:c-T_p2_D_-0_1_L24-120}}\hspace*{-5mm}
\subfigure{\includegraphics[scale=0.4,clip]{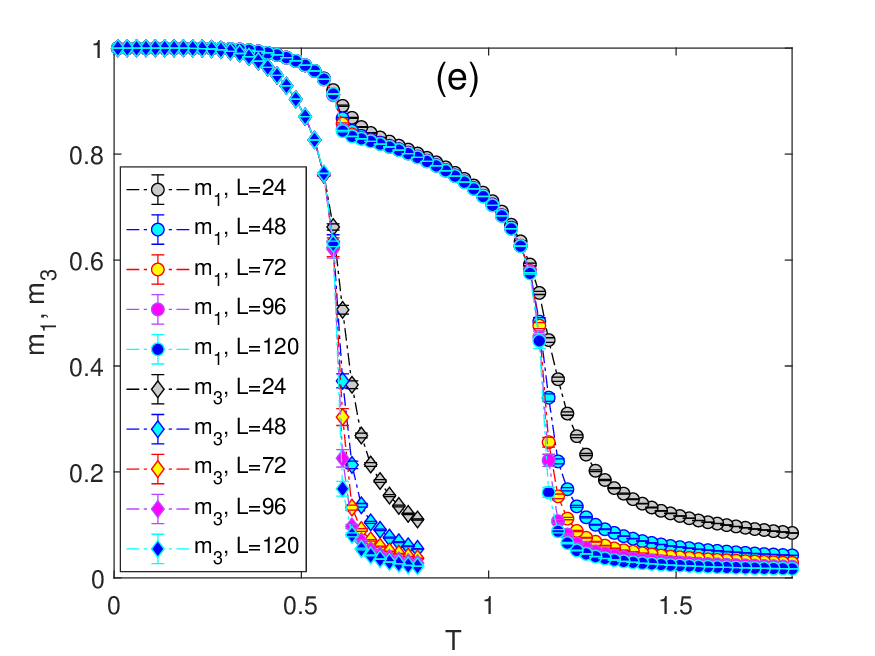}\label{fig:m-T_p2_D_-0_1_L24-120}}\hspace*{-5mm}
\subfigure{\includegraphics[scale=0.4,clip]{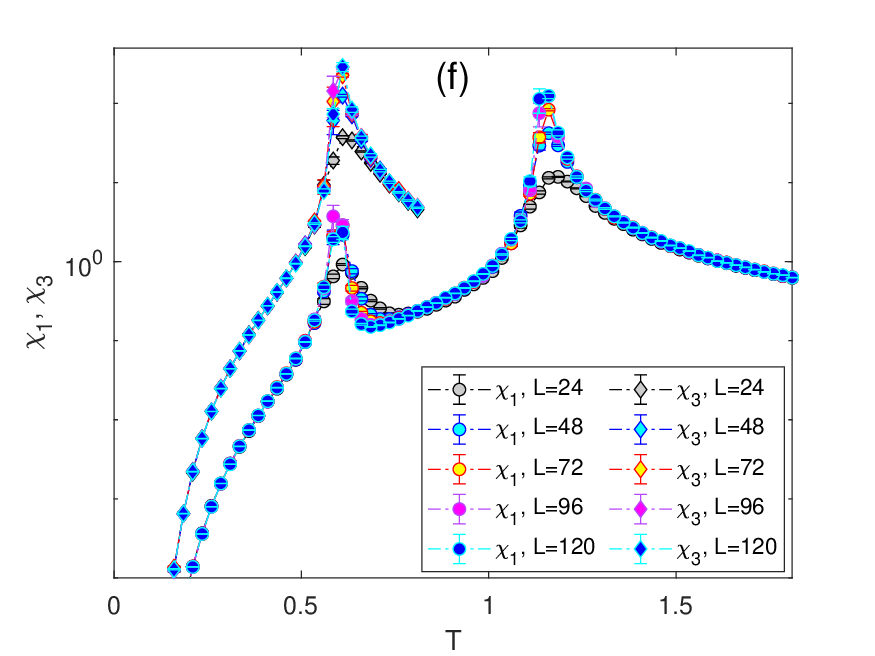}\label{fig:chi-T_p2_D_-0_1_L24-120}}\\ \vspace*{-3mm}
\subfigure{\includegraphics[scale=0.4,clip]{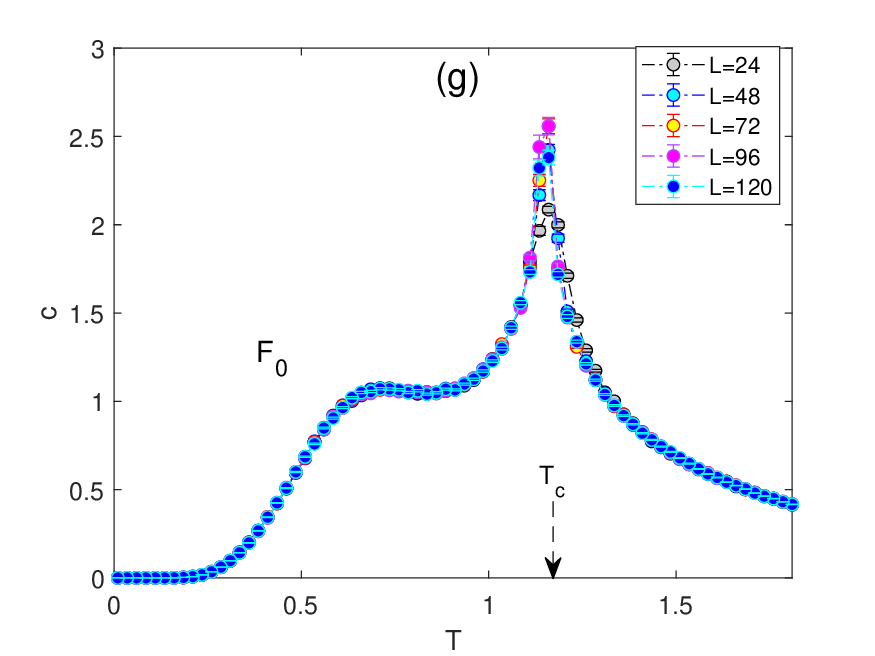}\label{fig:c-T_p2_D_0_1_L24-120}}\hspace*{-5mm}
\subfigure{\includegraphics[scale=0.4,clip]{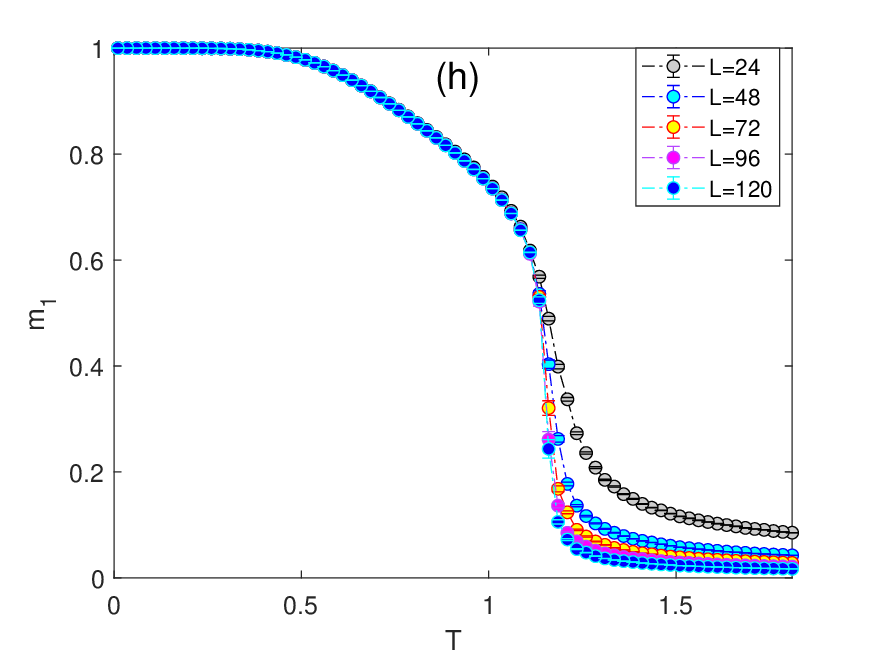}\label{fig:m-T_p2_D_0_1_L24-120}}\hspace*{-5mm}
\subfigure{\includegraphics[scale=0.4,clip]{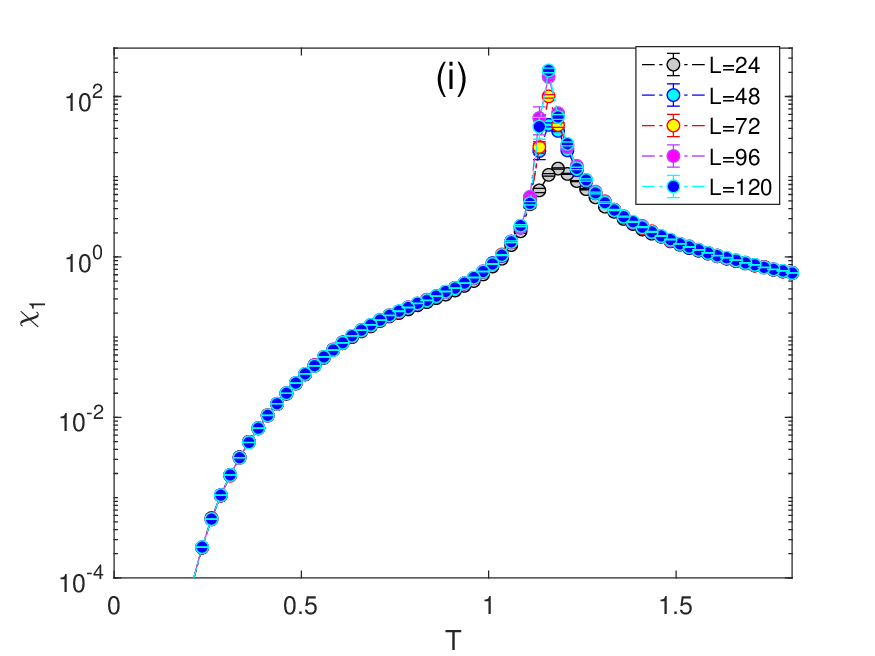}\label{fig:chi-T_p2_D_0_1_L24-120}}\\ \vspace*{-3mm}
\subfigure{\includegraphics[scale=0.4,clip]{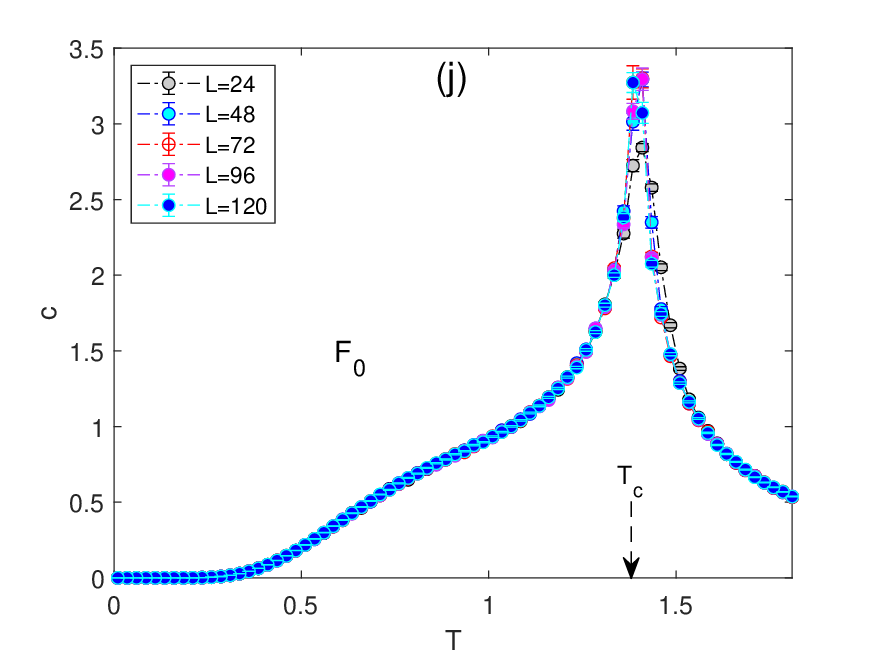}\label{fig:c-T_p2_D_0_5_L24-120}}\hspace*{-5mm}
\subfigure{\includegraphics[scale=0.4,clip]{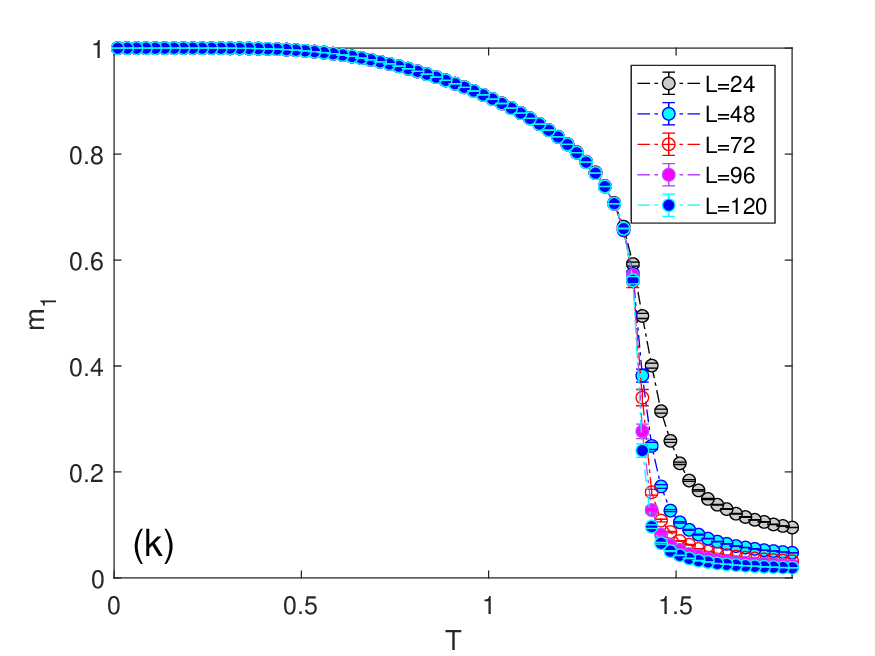}\label{fig:m-T_p2_D_0_5_L24-120}}\hspace*{-5mm}
\subfigure{\includegraphics[scale=0.4,clip]{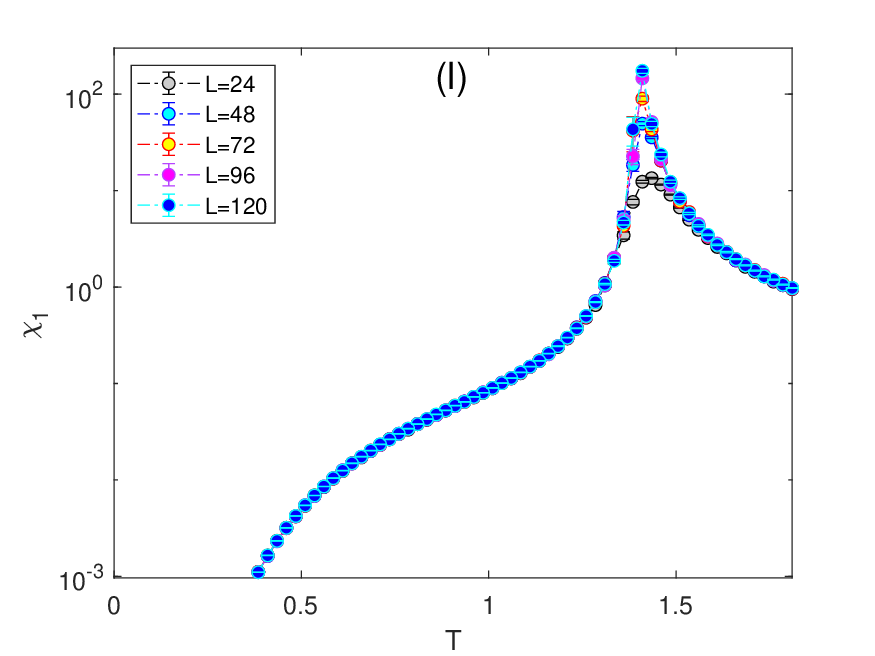}\label{fig:chi-T_p2_D_0_5_L24-120}}
\caption{Temperature dependencies of (a,d,g,j) the specific heat, (b,e,h,k) the magnetizations $m_1$ and $m_3$, and (c,f,i,l) the magnetic susceptibilities for $p=2$, (a-c) $h_2=-0.5$, (d-f) $h_2=-0.1$, (g-i) $h_2=0.1$, (j-l) $h_2=0.5$, and various lattice sizes. $T_{c1}$ and $T_{c2}$ are the critical temperatures at the respective $F_0-F_1$ and $F_1-P$ transitions for $h_2<0$ and $T_{c}$ denotes the critical temperature at the $F_0-P$ transition for $h_2>0$.}
\label{fig:x-T_p2_L24-120}
\end{figure}

In Fig.~\ref{fig:x-T_p2_L24-120}, temperature dependencies of (a,d,g,j) the specific heat, (b,e,h,k) the (generalized) magnetizations, and (c,f,i,l) the magnetic susceptibilities are presented for selected values of the field (a-c) $h_2=-0.5$, (d-f) $h_2=-0.1$, (g-i) $h_2=0.1$, (j-l) $h_2=0.5$, and different lattice sizes. These quantities and finite-size effects in there behavior will be helpful in understanding the nature of the ordering in the observed phases as well as the nature of the corresponding phase transitions. In particular, unlike in the $h_2=0$ case, one can observe higher sensitivity of the heights of the sharp specific heat peaks to the lattice size. Furthermore, there is no phase with the vanishing magnetization and divergent magnetic susceptibility that would suggest the presence of the BKT phase with QLRO. On the other hand, there are phases in which some of the generalized magnetizations vanish and some other retain finite values. Thus, different generalized magnetizations can be used as order parameters for different LRO phases.

In particular, for the cases with $h_2<0$ (the two upper rows), one can identify two LRO phases that can be described the order parameters $m_1$ and $m_3$. Figs.~\ref{fig:m-T_p2_D_-0_5_L24-120} and~\ref{fig:m-T_p2_D_-0_1_L24-120} show that the former decays at the transition temperature $T_{c2}$, that separates a kind of LRO ferromagnetic phase $F_1$ from the disordered paramagnetic phase P. The $F_1$ phase can be characterized as a mixed phase with the coexisting domains with the spins in the states $1$ and $2$ or $4$ and $5$ (see the snapshot). Since these states are in the same half-plane, there is a ferromagnetic LRO. As the temperature is lowered, one of the coexisting states is picked and the system undergoes the second phase transition at $T_{c1} < T_{c2}$ from the $F_1$ phase to the standard ferromagnetic phase $F_0$, in which all the spins align in the same direction. 

\begin{figure}[t!]
\centering
\includegraphics[scale=0.7,clip]{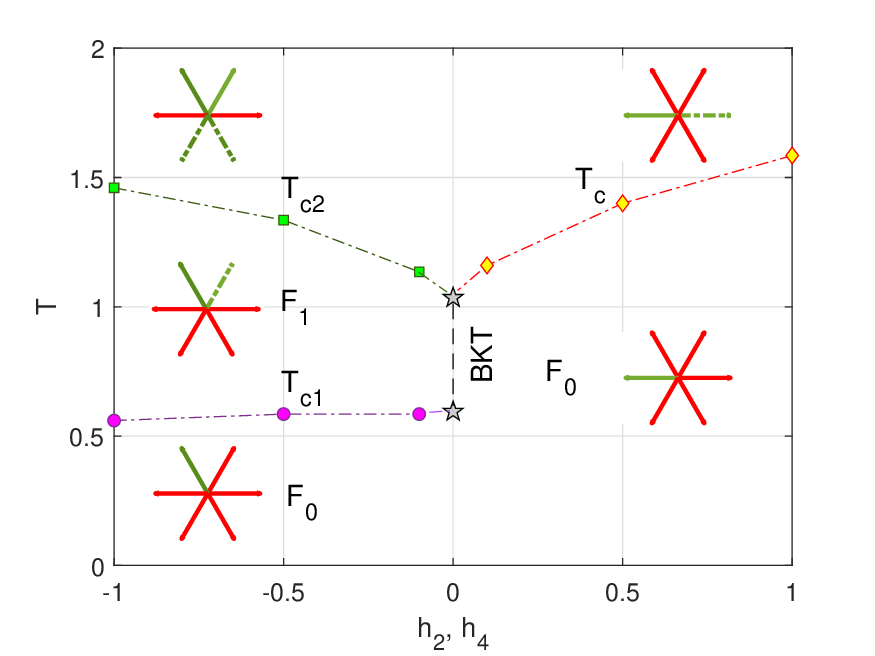}
\caption{Phase diagram for $p=2$ and $4$. $F_0$ and $F_1$ denote different FM LRO phases and BKT denotes the line of BKT transition points between $T_{c1}$ and $T_{c2}$ (marked by pentagrams) for $h_i=0$, $i=2,4$. In the schematic illustration of the characteristic spin states the green color represents the states selected by the symmetry-breaking field and thermal fluctuations with the dashed lines showing degenerate states.}
\label{fig:PD_p2}
\end{figure}

For $h_2>0$ (the two lower rows) the spins pick one from the two states preferred by the field $h_2$ and the system undergoes at $T_c$ a single phase transition to the FM phase $F_0$. The resulting phase diagram is presented in Fig.~\ref{fig:PD_p2}. As discussed above, the same phase diagram can also be obtained for $p=4$. We also numerically verified that the mean values of all the evaluated thermodynamic quantities as well as the transition temperatures and the critical exponents obtained for $p=4$ coincide with those for $p=2$ within the error bars. 

\begin{figure}[t!]
\centering
\subfigure{\includegraphics[scale=0.4,clip]{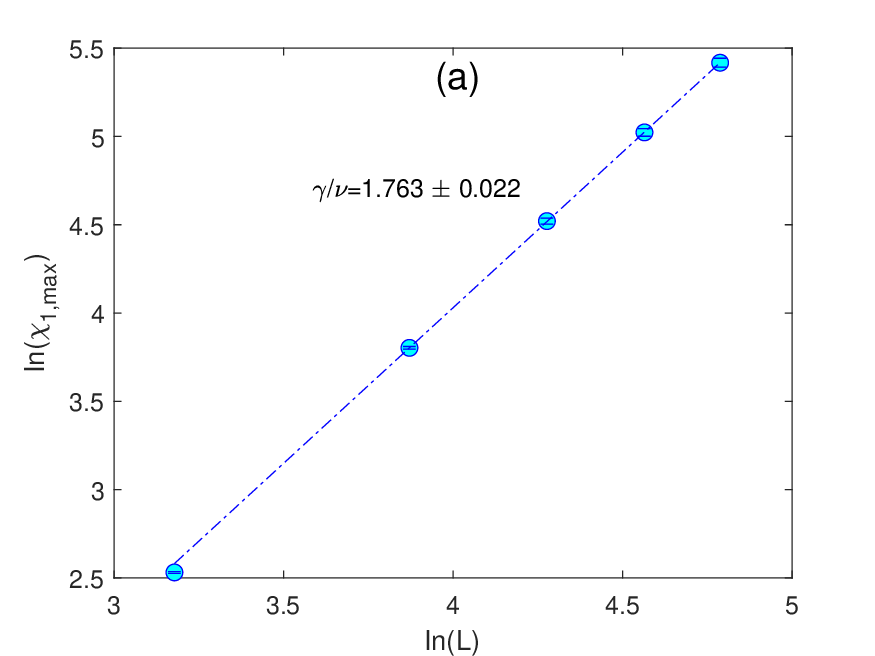}\label{fig:Tc2_fss_p2_xi_D-0_5}}\hspace*{-5mm}
\subfigure{\includegraphics[scale=0.4,clip]{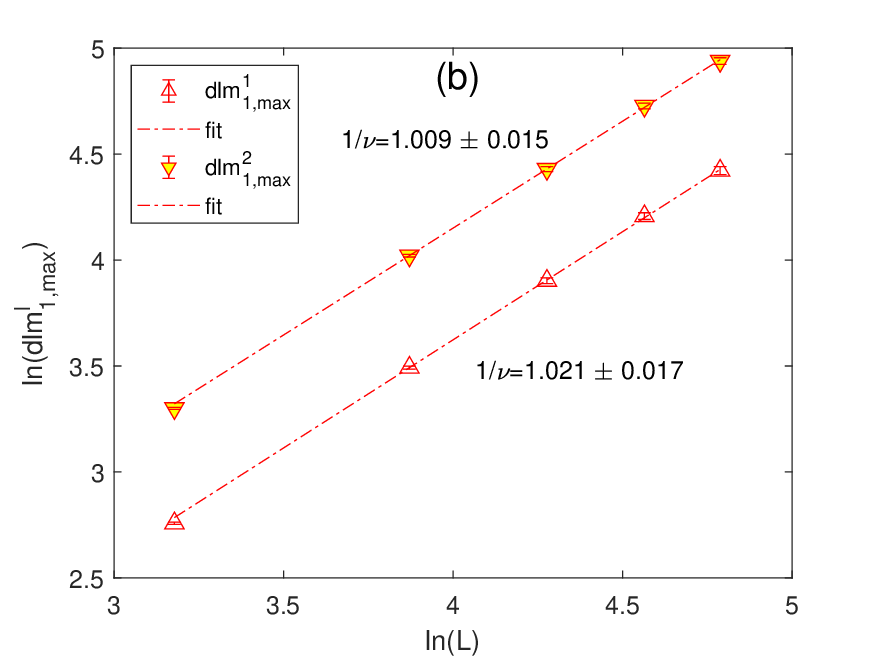}\label{fig:Tc2_fss_p2_dlm_D-0_5}}\hspace*{-5mm}
\subfigure{\includegraphics[scale=0.4,clip]{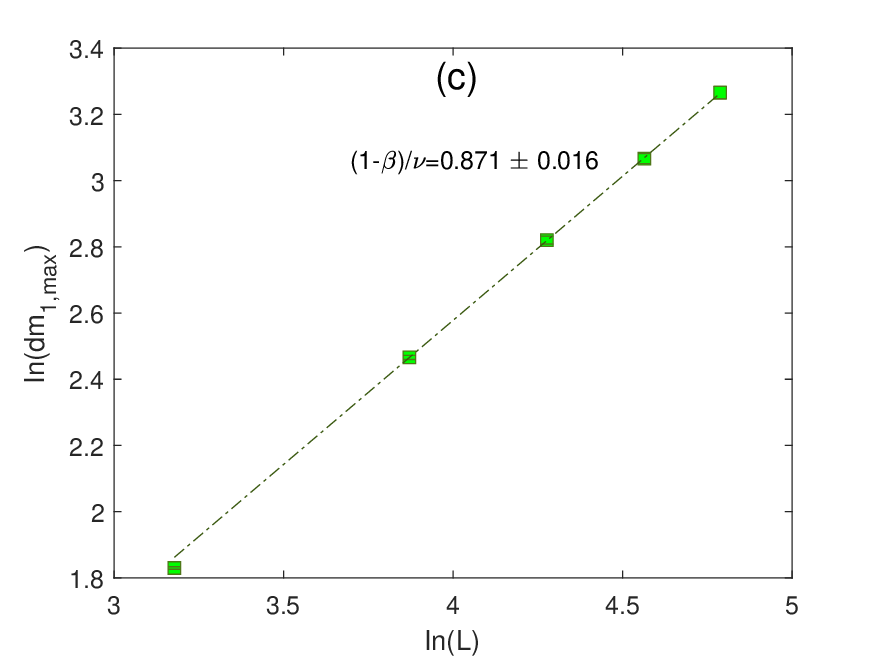}\label{fig:Tc2_fss_p2_dm_D-0_5}}\\
\subfigure{\includegraphics[scale=0.4,clip]{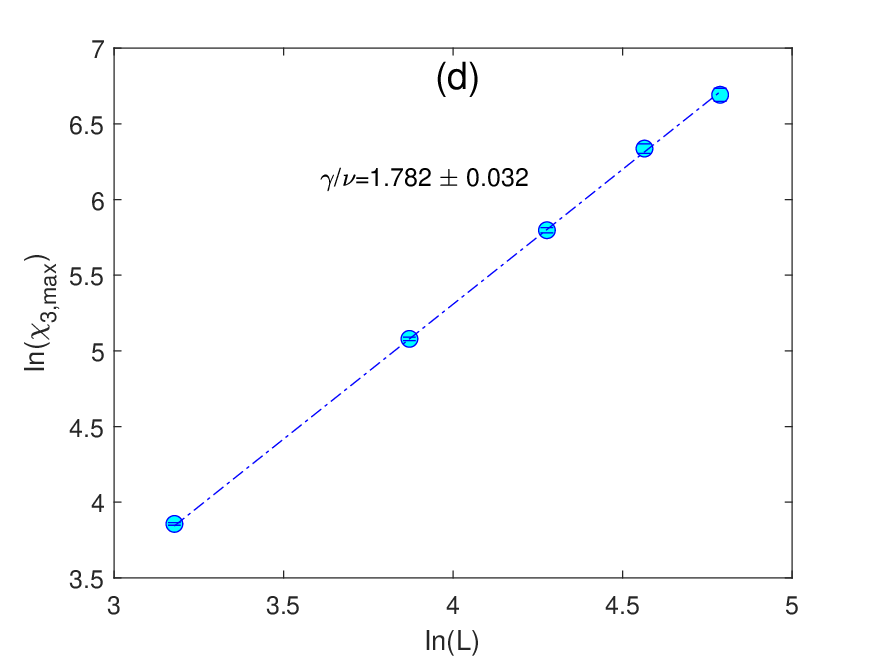}\label{fig:Tc1_fss_p2_xi3_D-0_5}}\hspace*{-5mm}
\subfigure{\includegraphics[scale=0.4,clip]{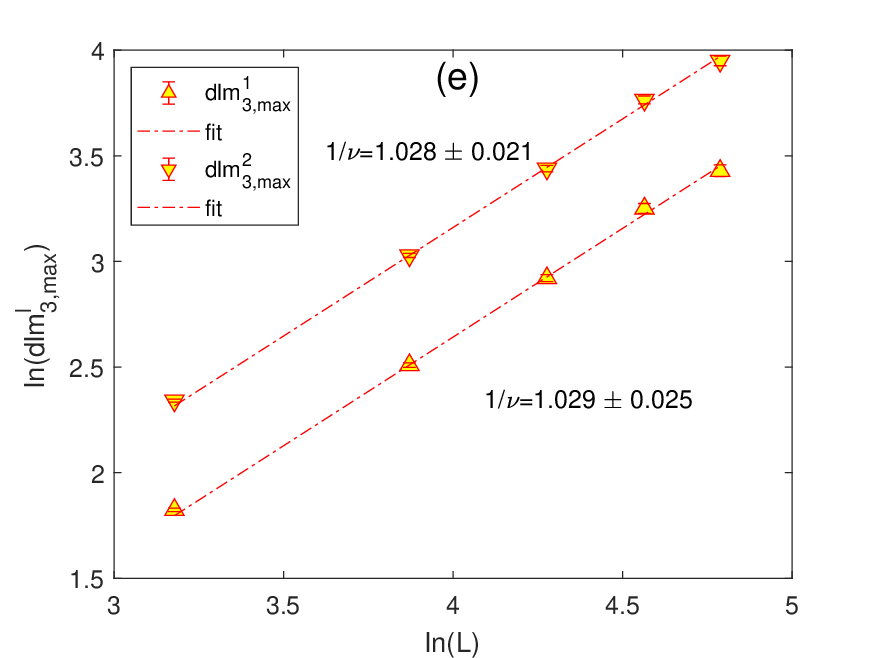}\label{fig:Tc1_fss_p2_dlm3_D-0_5}}\hspace*{-5mm}
\subfigure{\includegraphics[scale=0.4,clip]{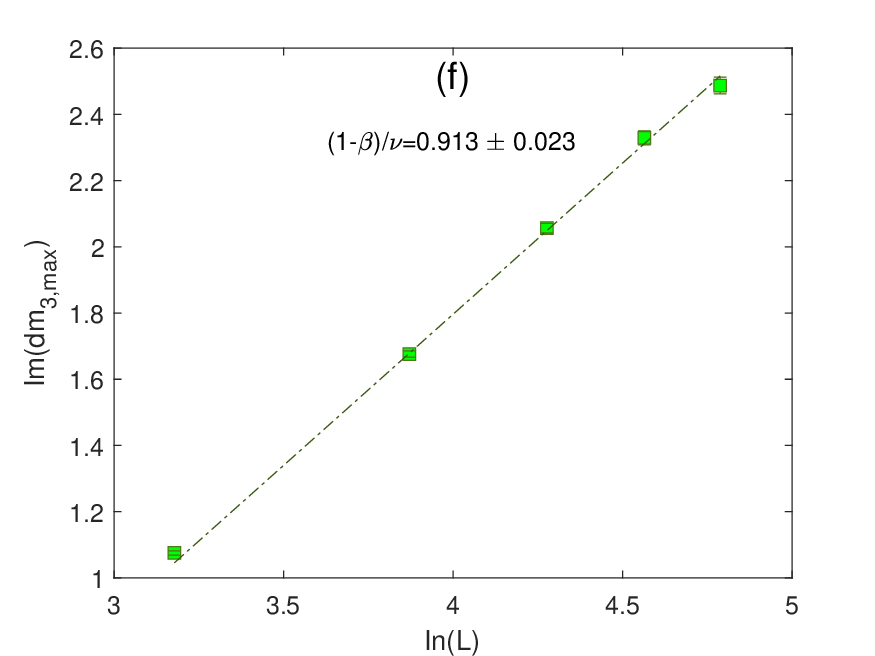}\label{fig:Tc1_fss_p2_dm3_D-0_5}}\\
\subfigure{\includegraphics[scale=0.4,clip]{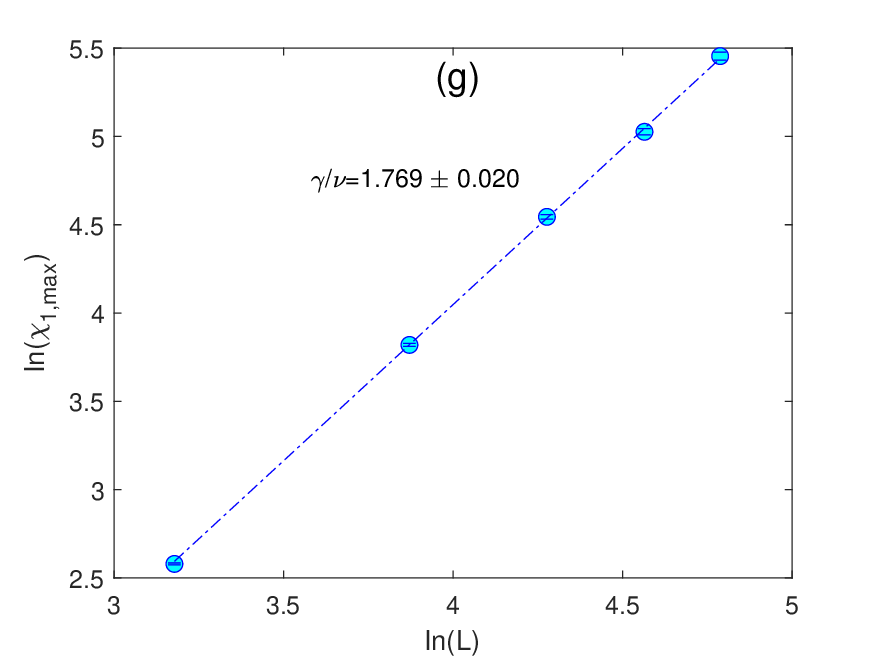}\label{fig:Tc_fss_p2_xi_D0_5}}\hspace*{-5mm}
\subfigure{\includegraphics[scale=0.4,clip]{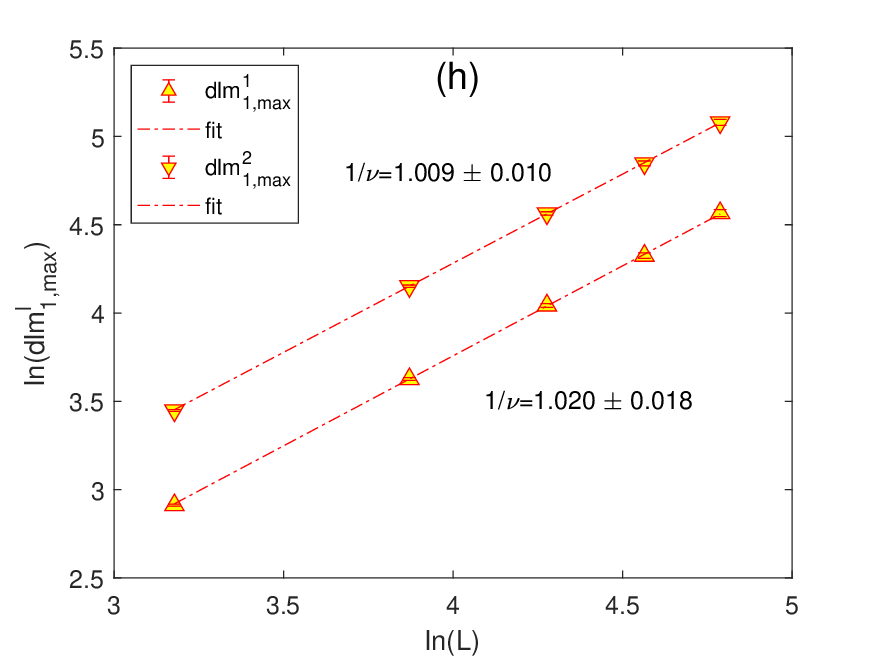}\label{fig:Tc_fss_p2_dlm_D0_5}}\hspace*{-5mm}
\subfigure{\includegraphics[scale=0.4,clip]{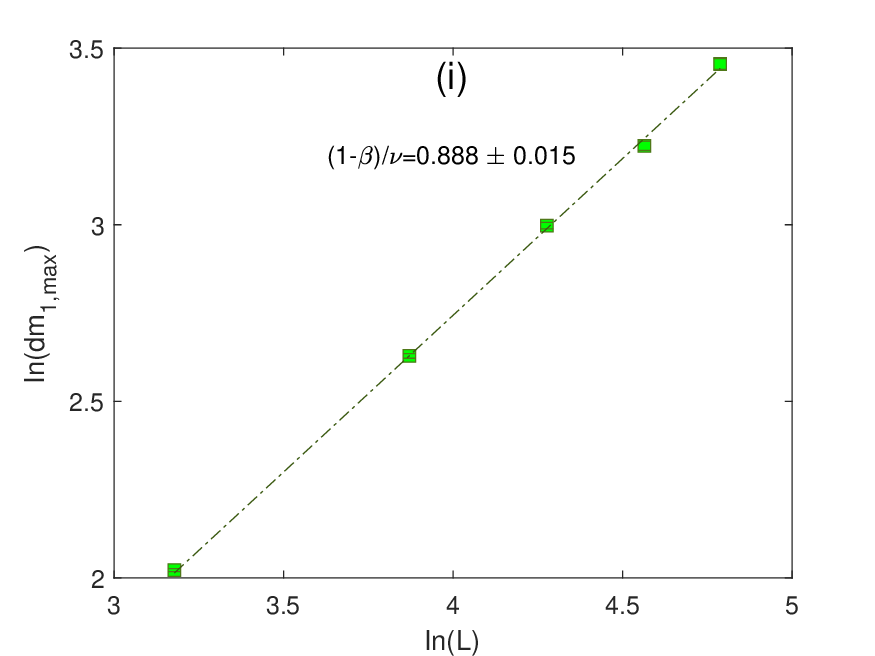}\label{fig:Tc_fss_p2_dm_D0_5}}
\caption{FSS analysis for $p=2$ of the quantities (a-c) $\chi_1$, $dlm^l_1$, $l=1,2$, and $dm_1$ at the transition temperature $T_{c2}$, (d-f) $\chi_3$, $dlm^l_3$, $l=1,2$, and $dm_3$ at the transition temperature $T_{c1}$, for $h_2=-0.5$ and (g-i) $\chi_1$, $dlm^l_1$, $l=1,2$, and $dm_1$ at the transition temperature $T_{c}$, for $h_2=0.5$.}
\label{fig:Tc_fss_p2_Dpm0_5}
\end{figure}

Next, by applying the FFS analysis in Fig.~\ref{fig:Tc_fss_p2_Dpm0_5} we estimate the critical exponents ratios at different phase transitions. In particular, the top row panels show the values of (a) $\gamma/\nu$, (b) $1/\nu$ and (c) $(1-\beta)/\nu$ obtained from the FSS of the quantities $\chi_1$, $dlm^l_1$, $l=1,2$, and $dm_1$, at the transition temperature $T_{c2}$, for $h_2=-0.5$. Similar analysis of the quantities $\chi_3$, $dlm^l_3$, $l=1,2$, and $dm_3$, shown in the middle row panels produces the critical exponents ratios at the transition temperature $T_{c1}$. The last three panels show the results at the single transition temperature $T_c$, for $h_2=0.5$. One can notice that in all the three cases the obtained critical exponents ratios comply with the Ising values $\gamma/\nu = 7/4$, $1/\nu = 1$, $(1-\beta)/\nu = 7/8$. We also verified (not shown) that the specific heat shows a logarithmic divergence as it should be at the Ising transition. The FSS results indicate that the asymptotic critical behavior is governed by the expected Ising fixed point, despite the presence of competing anisotropies and the proximity of the BKT regime. Nevertheless, the number of phase transitions with absence or presence of intermediate phases and the direct replacement of the BKT phase by a single transition are nontrivial consequences of the competition between anisotropies.

\subsubsection{Case $p=3$}
\begin{figure}[t!]
\centering
\includegraphics[scale=0.5,clip]{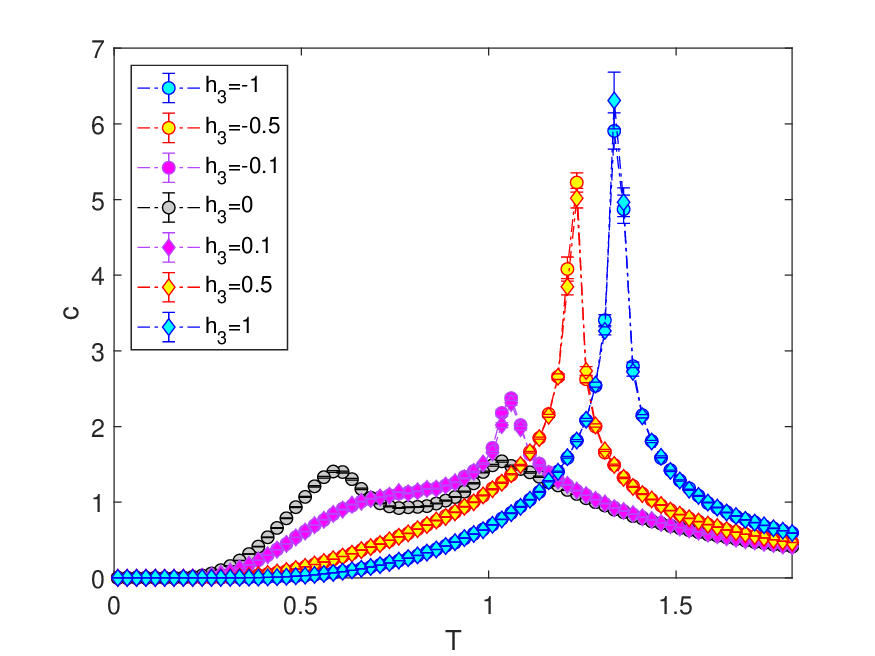}
\caption{Temperature dependencies of the specific heat for $p=3$, $L=72$, and various values of $h_3$.}
\label{fig:c-T_p3_D_all_L72}
\end{figure}

For $p=3$, the crystalline field selects three preferred orientations. For any finite value of $h_3$ the system exhibits a single phase transition from the disordered phase to an ordered phase with threefold symmetry. Indeed, the specific heat curves in Fig.~\ref{fig:c-T_p3_D_all_L72} suggest the same thermodynamic behavior for a given magnitude of the field $h_3$, regardless of its sign. Namely, for any $h_3 \neq 0$ the curves include one sharp peak that evolves from the high-temperature peak for $h_3=0$ and a hump that evolves from the low-temperature peak and gradually vanishes with the increasing $|h_3|$.

\begin{figure}[t!]
\centering
\subfigure{\includegraphics[scale=0.4,clip]{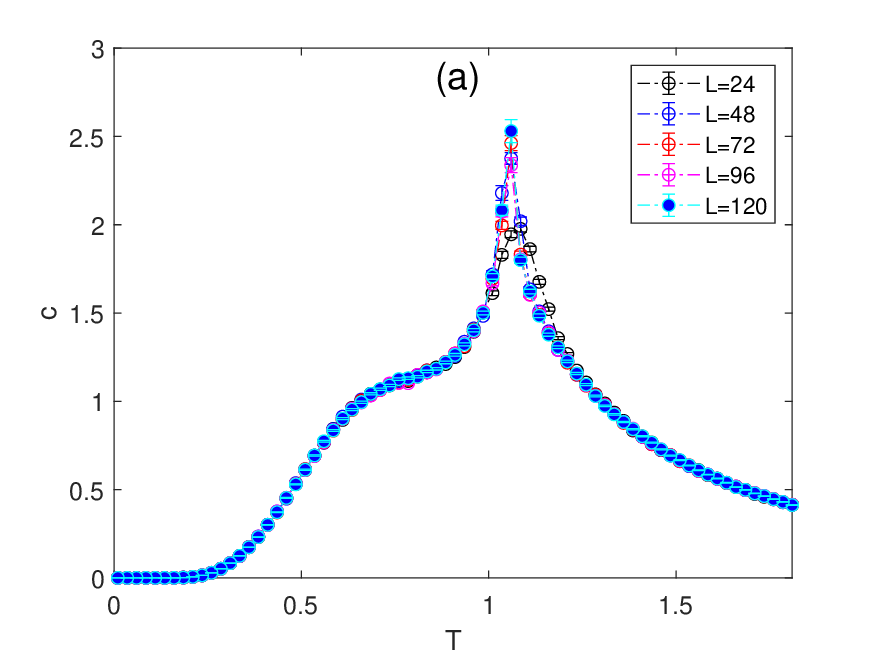}\label{fig:c-T_p3_D_-0_1_L24-120}}\hspace*{-5mm}
\subfigure{\includegraphics[scale=0.4,clip]{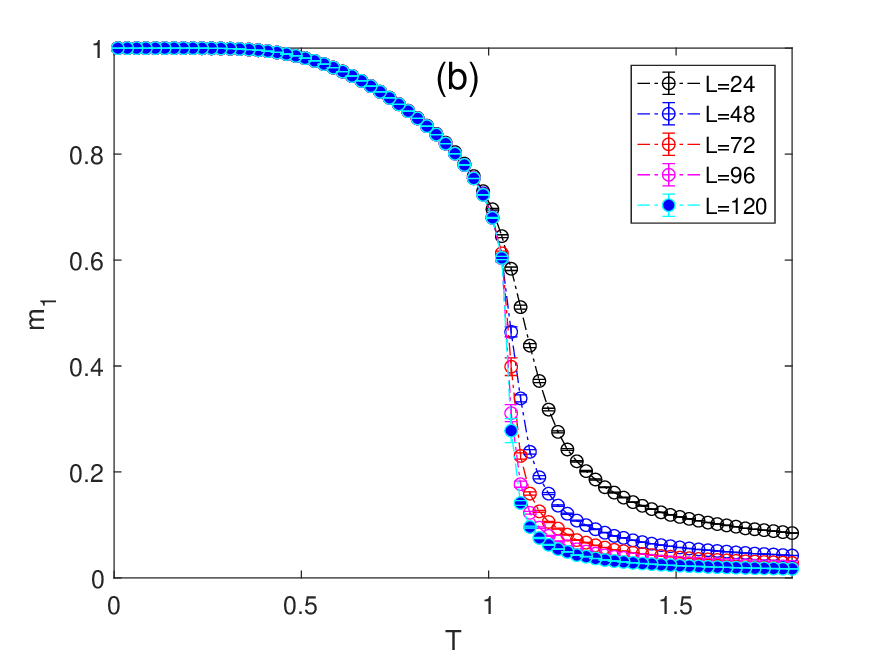}\label{fig:m-T_p3_D_-0_1_L24-120}}\hspace*{-5mm}
\subfigure{\includegraphics[scale=0.4,clip]{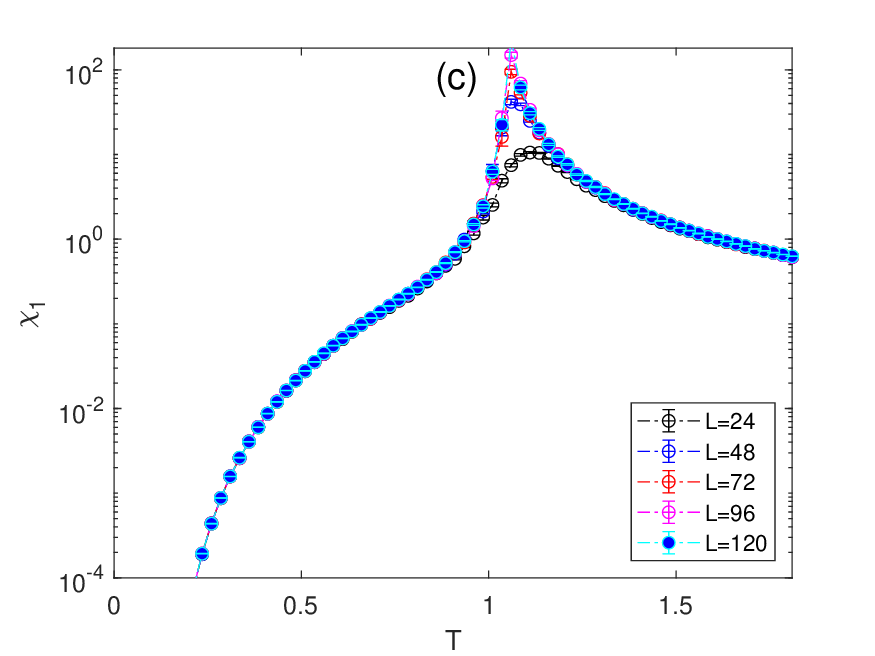}\label{fig:chi-T_p3_D_-0_1_L24-120}}
\subfigure{\includegraphics[scale=0.4,clip]{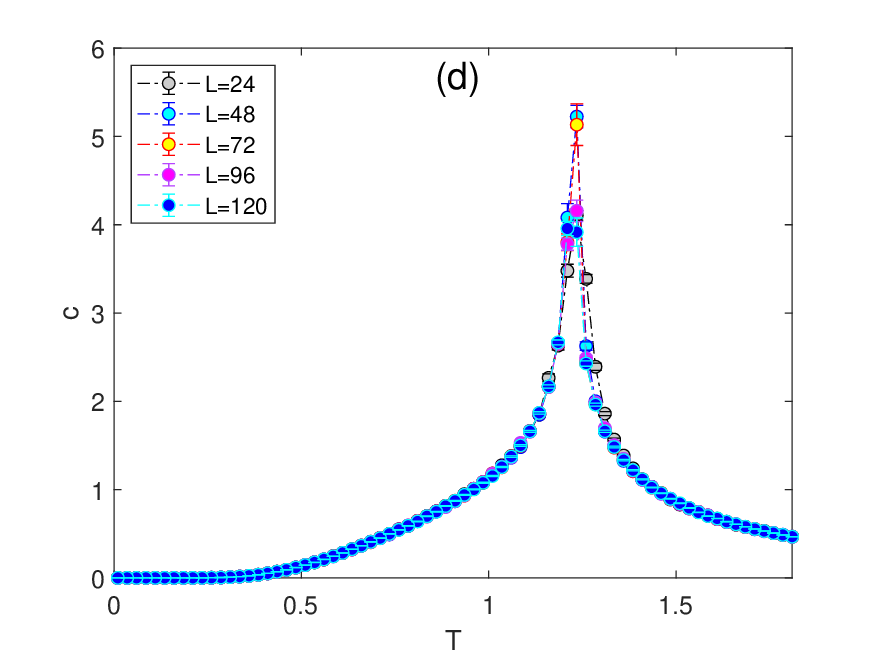}\label{fig:c-T_p3_D_-0_5_L24-120}}\hspace*{-5mm}
\subfigure{\includegraphics[scale=0.4,clip]{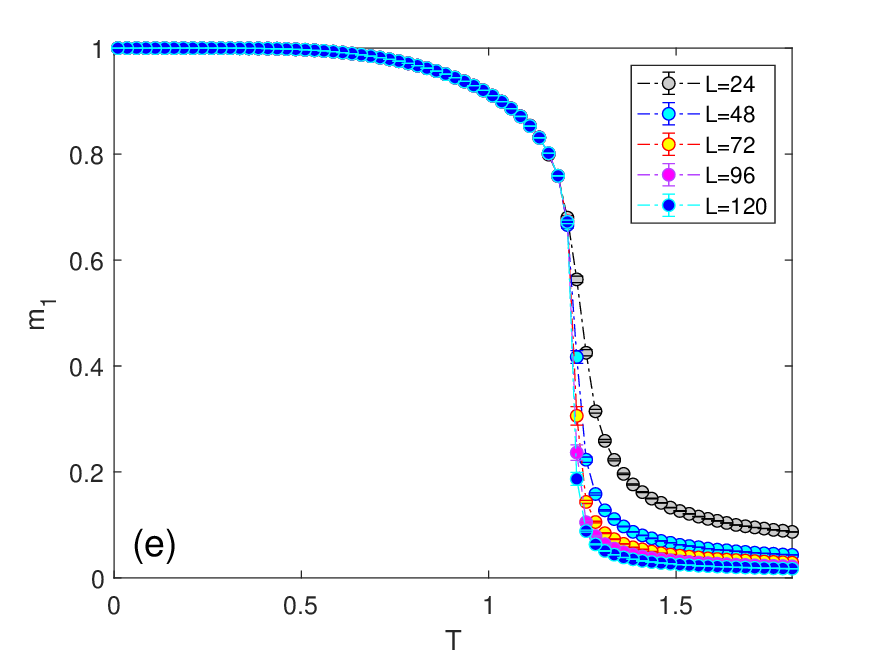}\label{fig:m-T_p3_D_-0_5_L24-120}}\hspace*{-5mm}
\subfigure{\includegraphics[scale=0.4,clip]{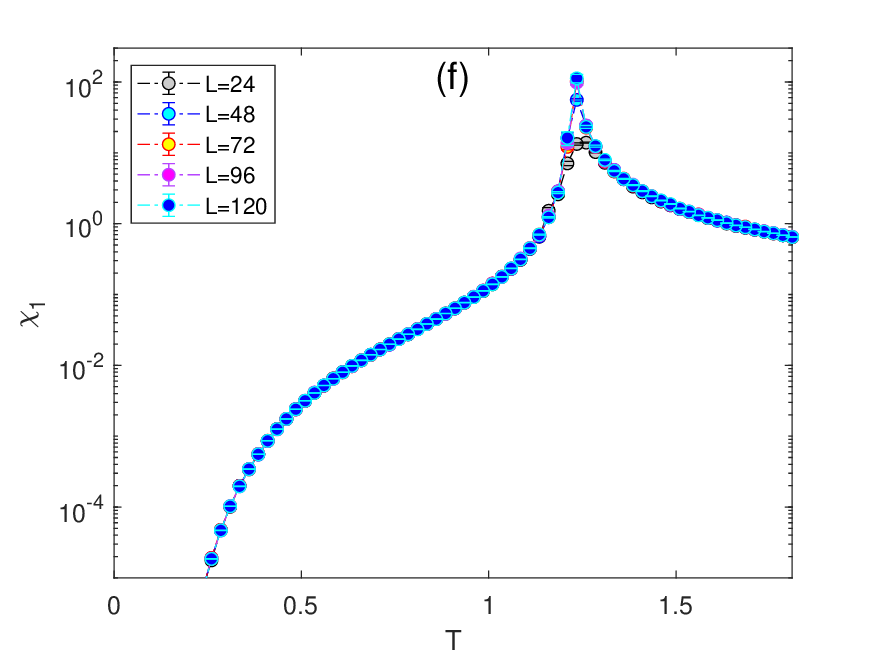}\label{fig:chi-T_p3_D_-0_5_L24-120}}
\caption{Temperature dependencies of (a,d) the specific heat, (b,e) the magnetization, and (c,f) the magnetic susceptibilities for $p=3$, (a-c) $h_3=-0.1$, (d-f) $h_3=-0.5$, and various lattice sizes.}
\label{fig:x-T_p3_L24-120}
\end{figure}

\begin{figure}[t!]
\centering
\includegraphics[scale=0.7,clip]{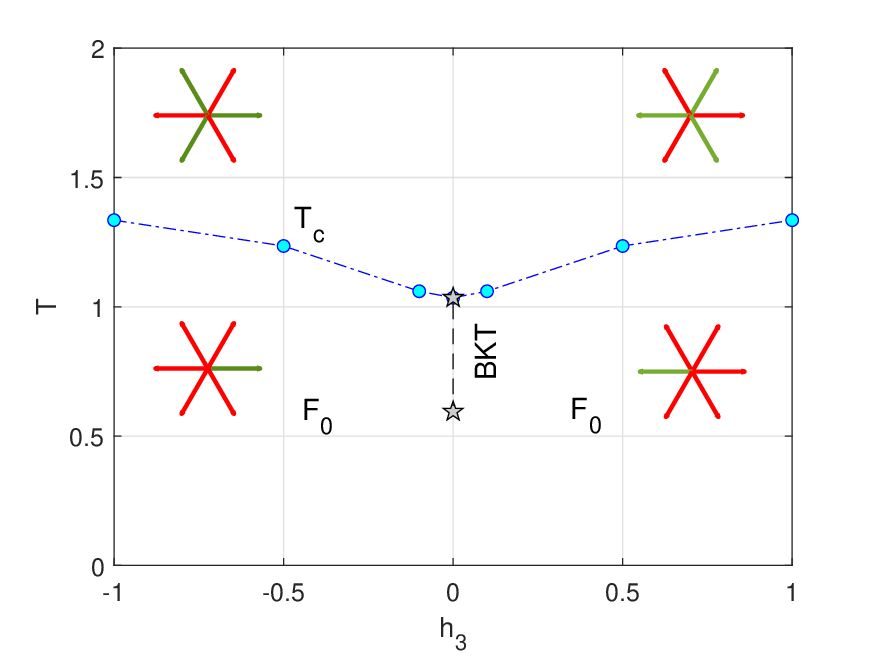}
\caption{Phase diagram for $p=3$. $F_0$ denotes the FM LRO phase and BKT denotes the line of BKT transition points between $T_{c1}$ and $T_{c2}$ (marked by pentagrams) for $h_3=0$. In the schematic illustration of the characteristic spin states the green color represents the states selected by the symmetry-breaking field and thermal fluctuations.}
\label{fig:PD_p3}
\end{figure}

Due to the insensitivity of the thermodynamic properties to the sign of the crystalline field the finite-size effects are presented in Fig.~\ref{fig:x-T_p3_L24-120} only for $h_3<0$. In all the quantities they only appear at a single transition point, which is accompanied with sharp peaks in both response functions, suggesting a non-BKT transition. Consequently, the resulting phase diagram (Fig.~\ref{fig:PD_p3}) consists of two wings symmetrically extending from the high-temperature BKT transition point at $h_3=0$ toward higher temperatures with the increasing $|h_3|$. The areas below the phase boundaries corresponds to the FM LRO phase $F_0$.

\begin{figure}[t!]
\centering
\subfigure{\includegraphics[scale=0.5,clip]{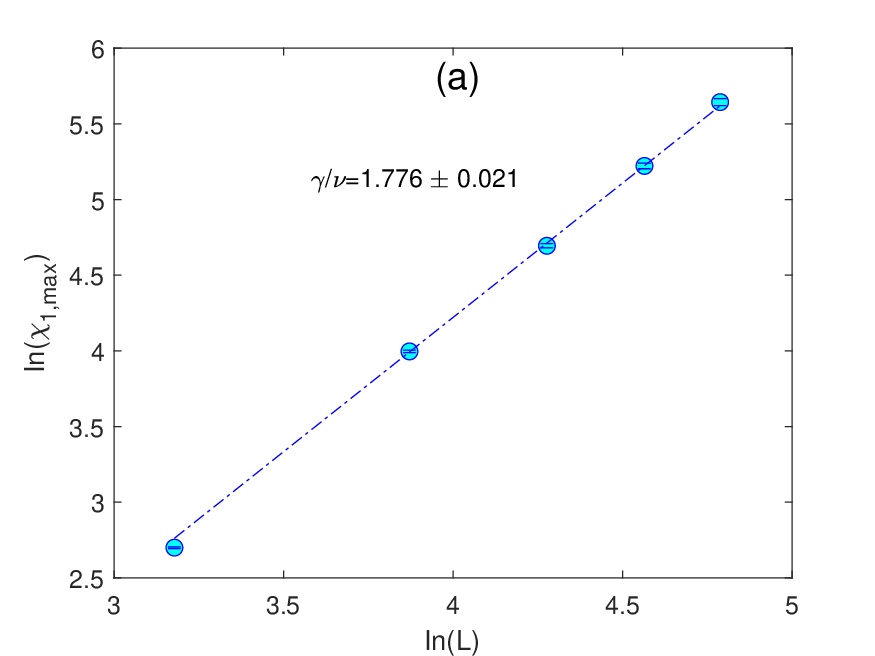}\label{fig:Tc_fss_p3_xi_D-0_5}}
\subfigure{\includegraphics[scale=0.5,clip]{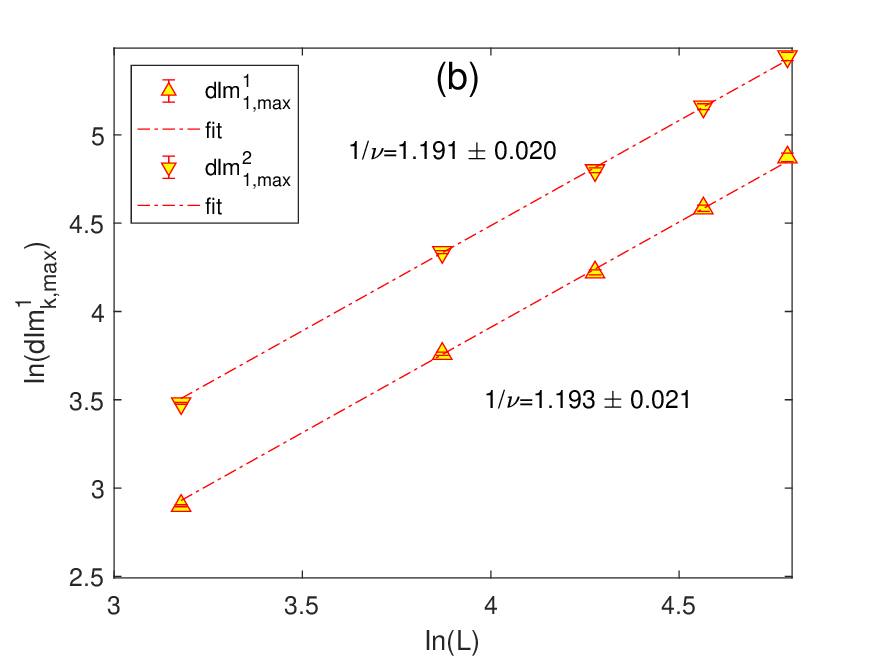}\label{fig:Tc_fss_p3_dlm_D-0_5}}\\
\subfigure{\includegraphics[scale=0.5,clip]{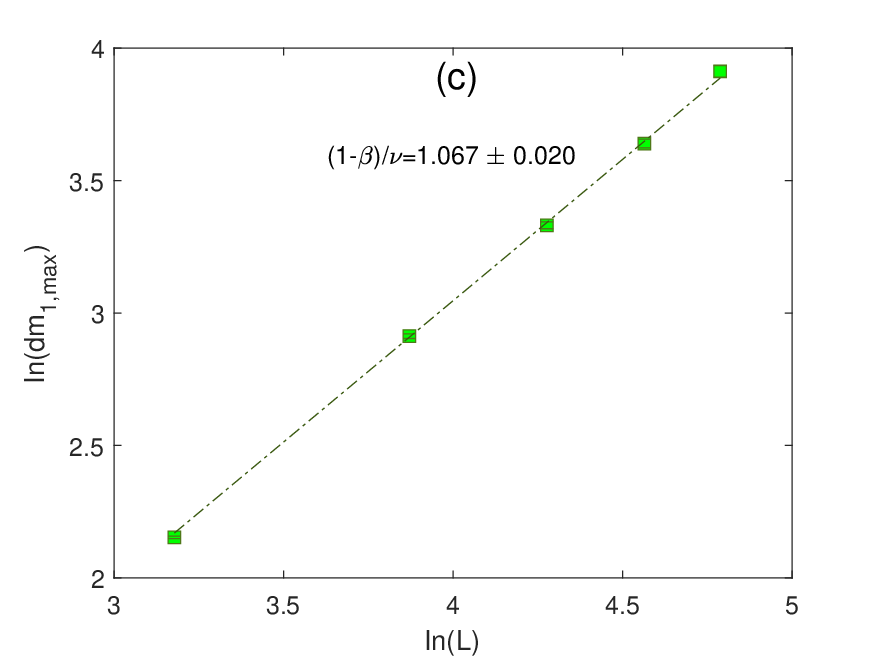}\label{fig:Tc_fss_p3_dm_D-0_5}}
\subfigure{\includegraphics[scale=0.5,clip]{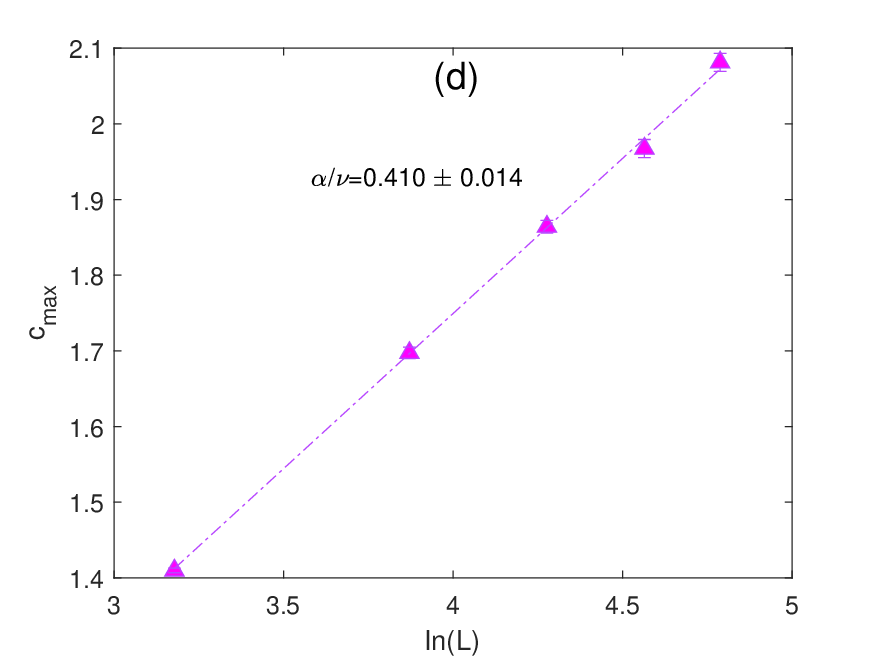}\label{fig:Tc_fss_p3_c_D-0_5}}
\caption{FSS analysis of the quantities (a) $\chi_1$, (b) $dlm^l_1$, $l=1,2$, (c) $dm_1$ and (d) $c$ at the transition temperature $T_{c}$, for $p=3$ and $h_3=-0.5$.}
\label{fig:Tc_fss_p3_D-0_5}
\end{figure}

The FSS analysis, performed for the representative value of $h_3=-0.5$, is presented in Fig.~\ref{fig:Tc_fss_p3_D-0_5}. The estimated values of the critical exponents ratios show that the transition is compliant with the three-state Potts universality with the expected values $\gamma/\nu = 26/15$, $1/\nu = 6/5$, $(1-\beta)/\nu = 16/15$ and $\alpha/\nu = 2/5$. As in the $p=2$ case, the transition replaces the BKT behavior of the pure clock model, indicating that the additional anisotropy is sufficient to suppress the critical phase.

Overall, these results demonstrate that the phase structure is governed by the competition between the intrinsic $\mathbb{Z}_q$ anisotropy and the externally imposed $\mathbb{Z}_p$ field. While symmetry considerations provide a useful starting point, they do not uniquely determine the number of transitions or the existence of intermediate phases. The Monte Carlo simulations reveal that even simple crystalline fields can qualitatively restructure the phase diagram by suppressing the BKT phase and inducing nontrivial ordering sequences.

\section{\label{sec:concl}Conclusion and Discussion}

In this work, we have investigated the two-dimensional $q$-state clock model in the presence of an additional $p$-fold symmetry-breaking crystalline field by means of Monte Carlo simulations. The main goal was to understand how competing discrete anisotropies modify the phase behavior of a system that, in its pure form, exhibits Berezinskii--Kosterlitz--Thouless (BKT) transitions and an intermediate critical phase.

Our results demonstrate that even relatively weak crystalline fields qualitatively restructure the phase diagram. In particular, the BKT behavior is replaced, within the accessible system sizes, by transitions consistent with conventional LRO. This finding is in accordance with the expectation that additional anisotropy terms act as relevant perturbations to the BKT fixed point, but the detailed form of the resulting phase diagram is not determined by symmetry alone and depends on the interplay between the intrinsic $\mathbb{Z}_q$ anisotropy and the externally imposed $\mathbb{Z}_p$ field.

For the representative cases with $q=6$ studied here, we find several distinct scenarios. For $p=2$, the behavior depends sensitively on the sign of the field. For $h_2>0$, the system undergoes a single transition from the disordered phase to a symmetry-broken ordered phase, consistent with Ising criticality. In contrast, for $h_2<0$ we observe a two-step ordering process with an intermediate phase characterized by partial symmetry breaking. The existence of two separate transitions, as well as the nature of the intermediate phase, reflects the competition between different subsets of energetically favored states and is not fixed by symmetry considerations alone. 

For $p=3$, the crystalline field selects three preferred orientations and the system exhibits a direct transition from the disordered phase to an ordered phase consistent with three-state Potts universality. In this case as well, the transition replaces the BKT behavior of the pure clock model, indicating that the additional anisotropy appears to control the long-distance behavior within the accessible length scales.

In the above studied cases the FSS analysis was only presented for the representative values of the field intensities of $h_p=-0.5$ and $0.5$. One may ask whether the Ising and Potts characters of the transitions for $p=2$ and $p=3$, respectively, persist along the entire phase boundaries, in particular in the proximity of the BKT transition line, i.e, for $h_p \to 0$. In fact, our FSS analysis at the respective transition temperatures in this region (for $h_2 = \pm 0.1$) produced somewhat larger deviations of the critical exponents ratios from the expected Ising values than for $h_2 = \pm 0.5$. Especially, the values of $\gamma/\nu$ exceeded $7/4$ by more than one standard deviation. For example, for $h_2=0.1$ the respective values at $T_c$ are: $\gamma/\nu = 1.822 \pm 0.023$, $1/\nu = 1.039 \pm 0.034$ (average from FSS of $dlm^1_1$ and $dlm^2_1$), and $(1-\beta)/\nu = 0.900 \pm 0.016$. However, dropping of the two smallest lattice sizes ($L=24$ and $48$) from the FSS fits resulted in the improved values $1.782 \pm 0.051$, $1.021 \pm 0.073$, and $0.875 \pm 0.036$ that already matched the Ising ones much better. Similar tendency was also observed for the $p=3$ case, for which the critical exponents ratios also showed larger deviations from the Potts values for $h_3=\pm 0.1$ than for $h_3=\pm 0.5$ but excluding of the smallest sizes from the FSS brought the values closer the expected Potts ones.

But what are the theoretical expectations regarding the fate of the BKT phase for $h_p \to 0$? From the renormalization-group perspective of Jos\'{e} \textit{et al.}~\cite{Jose1977}, the $p$-fold field corresponds to a perturbation whose relevance depends on its scaling dimension at the BKT fixed point. When relevant, this perturbation ultimately destroys the BKT phase in the thermodynamic limit. However, this destruction occurs via a crossover characterized by a rapidly growing correlation length as $h_p \to 0$. As a result, for small fields one expects an extended regime in which the system exhibits effective critical behavior on finite length scales, making it difficult to distinguish from a true BKT phase in numerical simulations. This explains the strong finite-size effects observed in our data near $h_p \to 0$. Although our numerical results are consistent with the instability of the BKT phase against finite crystalline fields, the crossover length scales become very large near $h_p \to 0$, making a definitive distinction between asymptotic and effective critical behavior numerically challenging.

In this study we focused on the $q=6$ case but the above considerations can be extended to other values of $q$. While a Table~\ref{table:energy}-type analysis can identify the energetically favored states for arbitrary $(p,q)$, it does not uniquely determine the phase structure. For example, for $q=8$ the pure model exhibits a well-developed BKT phase, and the introduction of a $p$-fold field leads to a competition between distinct periodicities. When $p$ divides $q$, the field selects a subset of clock states and may lead to multi-step ordering, while for incommensurate values of $p$ and $q$ additional frustration effects can arise. 

Taken together, the presented results show that the phase behavior of the model is governed by the competition between distinct locking mechanisms associated with the $\mathbb{Z}_q$ and $\mathbb{Z}_p$ symmetries. While symmetry arguments correctly identify the possible ordered states and their degeneracies, they do not uniquely determine the sequence of phase transitions, the existence of intermediate phases, or the fate of the BKT critical phase. These features emerge from the interplay of domain walls, vortex excitations, and entropic effects, and are captured here through numerical simulations.

From a broader perspective, the present model can be interpreted as the strong-anisotropy limit of generalized $XY$ models with multiple cosine terms, as considered in the renormalization-group analysis of Jos\'{e} \textit{et al.}~\cite{Jose1977}. In that framework, the competition between different harmonics leads to a rich structure of fixed points and crossover phenomena. Our results provide a complementary, fully discrete realization of this scenario, demonstrating how competing anisotropies reshape topological phase transitions and give rise to nontrivial ordering sequences.

Finally, the behavior observed here suggests several directions for further study. In particular, it would be interesting to explore systematically the dependence on $(p,q)$ and the role of commensurability, as well as to investigate the crossover regime near vanishing crystalline fields, where signatures of BKT physics may persist over intermediate length scales. Extensions to frustrated interactions or to systems with additional competing terms may reveal even richer phase diagrams. For example, higher-order harmonics in the $XY$ model can impose microscopic angular anisotropy and thus lead to abundant critical behavior~\cite{lee1985,carpenter1989,shi2011,hubscher2013,nui2018,poderoso2011,canova2016,zukovic2024,zukovic2025}. Such studies would further clarify the interplay between discrete symmetry breaking and topological excitations in two-dimensional systems.

\section*{Acknowledgment}
This work was supported by the Slovak Research and Development Agency under the Contract no. APVV-24-0091 and the Scientific Grant Agency of Ministry of Education of Slovak Republic under the Contract no. 1/0695/23.


\end{document}